\documentclass[graybox]{svmult}

\usepackage{amsmath}
\usepackage{type1cm}        
\usepackage{makeidx}         
\usepackage{graphicx}        
\usepackage{subcaption}                             
\usepackage{multicol}        
\usepackage[bottom]{footmisc}
\usepackage{bigstrut}
\usepackage{multirow}
\usepackage{newtxtext}       %
\usepackage[varvw]{newtxmath}       

\usepackage{float}

\def\0{{\bf 0}}
\def\x{{\bf x}}
\def\y{{\bf y}}

\newcommand{\n}{^{(n)}}

\newcommand{\F}{\mathbf{F}}
\newcommand{\Q}{\mathbf{Q}}

\newcommand{\X}{\mathbf{X}}
\newcommand{\Y}{\mathbf{Y}}
\newcommand{\Ind}[1]{I \left(#1\right)}
\newcommand{\rmE}{{\rm E}}

\def\pms{\mspace{-1mu}{\scriptscriptstyle\pm}}

\makeindex             

\begin{document}

\title*{Some novel aspects of quantile regression: local stationarity, random forests and optimal transportation}
  \titlerunning{Novel aspects of quantile regression} 
\author{Manon Felix, Davide La Vecchia, Hang Liu and Yiming Ma}
\institute{Manon Felix and Davide La Vecchia  \at Research Center for Statistics and Geneva School of Economics and Management, University of Geneva, \email{manon.felix@unige.ch and davide.lavecchia@unige.ch}
\and Hang Liu  \at  International Institute of Finance, School of Management, University of Science and Technology of China \email{hliu01@ustc.edu.cn}
\and Yiming Ma \at Department of Statistics and Finance, School of Management, University of Science and Technology of China  \email{mayiming@mail.ustc.edu.cn}}
%
%
\maketitle

\abstract{This paper is written for a Festschrift in honour of Professor Marc Hallin and it proposes some developments on quantile regression. We connect our  investigation to Marc's scientific production and we  present some theoretical and methodological advances for quantiles estimation in non standard settings.  We split our contributions in two parts. The first part is about conditional quantiles estimation for nonstationary time series. 
The second part is about conditional quantiles estimation  for the analysis of multivariate independent data in the presence of possibly large dimensional covariates. 
Monte Carlo studies illustrate numerically the performance of our methods and compare them to some extant techniques.}

\section{Introduction}
\label{sec:1}

A decision-theoretical approach to quantile-oriented and  rank-based  inference has been a  \textit{fil rouge}  running through Professor Marc Hallin’s entire scientific life. It provides coherence to his otherwise very broad, long, and diverse list of 
contributions to mathematical statistics and econometrics. Following the \textit{fil rouge}, we build on some Marc's  (old and recent)  research  outcomes: we propose some theoretical and methodological developments for conditional quantile estimation. Our investigation is motivated by the need for modeling and estimating conditional quantiles in two non standard and challenging settings: nonstationary time series  (section \ref{sec:2}) and multivariate independent data (section \ref{sec:3}). 
To introduce our contributions, we review the extant literature. Rather than embarking in a pointless attempt at being comprehensive, we decided to put the emphasis on the key results and on Marc's research outcomes  connected  to our developments.

\subsection{Quantile estimation for nonstationary AR processes}

Quantile estimation for time series and related autoregression rank scores have been discussed in many of Marc's publications. For instance, \cite{EH02} propose
estimators of the quantile density function associated with the innovation density of an autoregressive model of order $p$ (AR($p$)) and their estimators are based on 
autoregression quantiles. 
\cite{HJ99} construct locally asymptotically optimal tests based on autoregression rank scores, whose concept  
was introduced by \cite{GJ92} 
and further developed by \cite{KS95}
 in the univariate time series context. We refer also to  \cite{hallin2007serial} for the concept of  serial autoregression rank scores.
 
 The assumption of stationarity lies at the heart of all these developments. 
However, empirical studies suggest that  stationarity appears to be doubtful (to say the least). 
One possible approach to deal with this inference aspect is related to the concept of local stationarity; see \cite{D12} for a review. In that framework, the object of interest is a stochastic process whose parameters are changing smoothly over time, in such a way that it can be locally approximated by a stationary process. Importantly, by rescaling the observation period to the unit interval, estimators for the time-varying parameters can be  obtained using windowed (in time) estimating equations. This concept has been successfully applied to various types of processes, including AR processes \cite{D97}, ARCH processes \cite{FSSR08}, nonlinear AR processes \cite{V12},  scalar diffusion processes \cite{KL12}, multivariate diffusion processes \cite{DLV19}, and  Markov processes \cite{Tr19}. Some recent developments  for quantile spectral analysis are available in \cite{BVKDH17}, while \cite{XSZ22} (and  to some extent \cite{ZW09}) study conditional quantile estimation. 

 The approach that we adopt in this paper has a spirit similar to the one of
\cite{DLV19}, who explain how to conduct inference on the time-varying parameters of diffusions and how to derive the asymptotics of the corresponding estimators. 
However,  \cite{DLV19} 
 focus on the first two (infinitesimal) moments of the process, whilst here we study the problem of quantile regression and we look at the entire conditional distribution. To this end, we consider  time-varying autoregression  quantile  estimation for a process  that we observe over a time span $[0,n]$. 
The central idea of our estimation method relies on solving a sequence of localized in time quantile regression optimization problems, where the localization is achieved using a kernel with compact support.  
We make use of  the local polynomial quantile regression technique to estimate the model parameters and
we derive the asymptotic properties  of proposed estimators, studying their bias and their asymptotic distribution.  The resulting inference procedure 
 complements the results already available in \cite{XSZ22}. 

\subsection{Nonparametric multiple-output center-outward quantile regression and random forests for independent data}

The topic of {multivariate nonparametric quantile regression} has been attracting the interests of the research community for a long time; see  \cite{HM17} for a review.
 Let us summarize its key inference issues. 
 
 The problem of quantile regression is well-understood for the case of univariate random variables. However, an extension to the multivariate case (also called multiple-output case) is not straightforward because most of the extant definitions of regression quantiles (namely, the traditional definition, the $L_1$ definition and  the definition based on regression quantile hyperplane) exploit the canonical ordering of the real line. Such an ordering no longer exists in $\mathbb{R}^d$, $d \geq 2$. This entails that notions like quantiles, check function, distribution function, signs, and ranks do not clearly extend to higher dimensions. Some solutions to this problem are already available, like the directional and direct approaches; see \cite{HM17} p.187-193. 

Making use of statistical concepts related to optimal transportation theory (see \cite{H22} and \cite{LVRI23}  for a discussion on the use of Monge-Kantorovich results in statistics),
\cite{dBSH22} define nested conditional center-outward quantile regression contours and regions, with given conditional probability content irrespective of the underlying 
distribution. Their  graphs constitute nested center-outward quantile regression tubes.  \cite{dBSH22} illustrate how to construct empirical counterparts of these population 
concepts, yielding interpretable empirical regions and contours. Their construction is based on two steps: in Step 1, one specifies a set of weights and uses them to construct an 
empirical distribution of the multivariate response variable conditional on some values of the multivariate covariates;  in Step 2, one computes the corresponding empirical center-
outward quantile map, resorting on Monge-Kantorovich's results.

Building on that approach, we combine the theory of random forests with the novel concepts of center-outward quantiles. The proposed 
inference method merges some results rooted in the machine learning literature (random forests \cite{B01}, quantile regression forests \cite{MR06}, and generalized random forests \cite{ATW19}),  in mathematics (optimal transportation, \cite{V08}) 
and in statistics (multivariate quantile, \cite{Hallin2021}).  

At a high level, our procedure makes use of random forests as an adaptive neighbourhood classification tool, which we combine with the multiple-output center-outward quantile regression of  \cite{dBSH22}. We grow the trees mimicking the logic of standard random forests, obtaining, for every regressors value, a set of weights for the original multivariate response. This idea has the same spirit as \cite{MR06} but it deals with a multivariate response variable and needs a novel approach to specify and compute the weights. To this end, we build on the theory of multivariate random forests of 
\cite{SX11}:  we apply the resulting weights to estimate the conditional distribution, which is defined as the weighted distribution of observed response variables, as in Step 1 of \cite{dBSH22}. Then, moving along the same lines as Step 2 in \cite{dBSH22}, conditional center-outward quantile maps are obtained. This construction yields a novel inference tool which is able, by design, to alleviate the curse of dimensionality for the analysis of multi-output variables in the presence of large dimensional covariates.

%
%

\section{Local stationarity}\label{sec:2}

\subsection{A motivating example} \label{Mot1}

To illustrate the impact that ignoring the nonstationarity of a time series may have on the estimation of the conditional quantiles (and hence on the whole conditional distribution), let us consider the following simple motivating example.  To begin with, let us fix the notation and consider the AR(1) process having dynamics
\newcommand{\fin}{i/n}
\newcommand{\fiin}{\frac{i-1}{n}}
 $X_i = \phi_1 X_{i-1} + e_i$,
where $\{e_i, i=1,...,n\}$ are i.i.d., zero mean and unit variance, with cumulative distribution function (cdf) $F$ and $X_0=0$.  Let $\mathcal{F}_{i-1}$ be the $\sigma$-field containing the
 past values of the process. The theoretical conditional quantile, at a probability $\tau\in(0,1)$, of $X_i$ given $\mathcal{F}_{i-1}$ is 
 \begin{equation}
 Q_\tau (X_i  \vert \mathcal{F}_{i-1}) = \phi_1 X_{i-1} + F^{-1}(\tau)  = \phi_1X_{i-1} + \alpha (\tau) = \boldsymbol{U}_i^{\top} \boldsymbol{\theta}(\tau) ,
 \label{QAR_stat_case}
 \end{equation}
with $\boldsymbol{\theta}(\tau) = (\alpha(\tau), \phi_1)^{\top}$ and $\boldsymbol{U}_i = (1, X_{i-1})^{\top}$. To estimate the model parameters we use the standard quantile regression approach and obtain
\begin{equation}
\hat{\boldsymbol{\theta}}(\tau)=\underset{\boldsymbol{\theta} \in \Theta}{\operatorname{argmin}} \sum_{i = 1}^n \rho_\tau\left(X_i- \boldsymbol{U}_i^{\top} \boldsymbol{\theta}(\tau) \right),
\label{Eq: est_constQAR}
\end{equation}
where  $\rho_\tau$ is the check-function defined as $\rho_\tau(u)=u(\tau-I(u<0))$ and $I(\cdot)$ is the indicator function.  A simple plug-in of $\hat{\boldsymbol{\theta}}(\tau)$ (called the autoregression quantile) into the expression of $Q_\tau( X_i | \mathcal{F}_{i-1})$ yields $\hat Q_\tau (X_i | \mathcal{F}_{i-1})= \boldsymbol{U}_i^{\top} \hat{\boldsymbol{\theta}}(\tau)$, which is an estimate of the conditional quantile.

\begin{figure}
\begin{center}
 \includegraphics[width = 0.85\textwidth, height = 0.285\textheight]{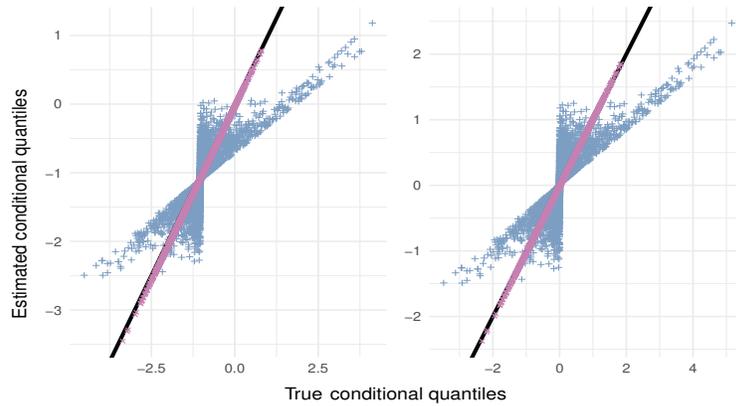}
 \end{center}
 \caption{Conditional quantiles (true versus estimated)  scatter plots for an AR(1). Left plot: probability level $0.15$; Right plot: probability level $0.5$. 
 In each plot: the (blue) $+$'s are for the case of time-varying  parameter ; the (pink) $\times$'s are for the case of time constant parameter; the bold continuous line is for the 45 degrees line.}
\label{Fig.1}
 \end{figure}

Now assume that the underlying AR(1) has a time-varying parameter 
$\phi_1(i/n)$. 
The resulting process is nonstationary, with non constant true conditional quantiles (and more generally, with time-changing conditional density) having expression as
in (\ref{QAR_stat_case}), 
with the time constant $\boldsymbol{\theta}(\tau)$ replaced by the time-varying  $\boldsymbol{\theta}(i/n \mid \tau) = (\alpha(\tau), \phi_1(i/n))^{\top}$. Assume that we conduct inference ignoring the fact that the  process is nonstationary:  
we estimate $\boldsymbol{\theta}(\tau)$ using (\ref{Eq: est_constQAR}) and we define the conditional quantile via the plug-in of resulting estimates into the expression of $Q_\tau (X_i | \mathcal{F}_{i-1})$. 
It is easy to conjecture that the time varying nature of the model parameter entails that the estimated conditional quantiles 
are not reliable estimates of the true time-varying conditional quantiles. 
To illustrate numerically this aspect, we conduct a Monte Carlo study. 
 Specifically, we consider a stationary process with parameter $\phi_1=0.5$ and a nonstationary 
one with time-varying parameter  $\phi_1({i}/{n})= c ({i}/{n}) + d ({i}/{n})^{2.5}$,  where we set $c = 0.1$, $d = 0.85$, $n=4000$ and the innovations are i.i.d. with standard Gaussian distribution. 
In Figure \ref{Fig.1}, we display  the estimated and true conditional quantiles via scatter plots, for the stationary and nonstationary case, in one Monte Carlo run. For both $\tau=0.15$ (left plot) and $\tau=0.5$ (right plot), when the underlying process is stationary, the true and the estimates conditional quantiles are very similar (the $\times$ symbols are overlapping with the 45 degrees line). In contrast, when the underlying process is nonstationary, the estimated conditional quantiles are biased (the  $+$ symbols are scattered around the  45 degrees line). This bias is due to the fact that the estimation procedure in (\ref{Eq: est_constQAR}) does not take into account the time-varying nature of the process. 
In the next subsection we explain how to derive a class of estimators, which is able, by design, to cope with this issue. 

\subsection{Autoregression quantile estimation} \label{LS_theory}

\subsubsection{Stationary case}

\textit{Key notions.} 
Let us consider a stationary AR($p$), $p \in \mathbb{N}$:
$X_i=\phi_0+\phi_1 X_{i-1}+\cdots+\phi_p X_{i-p}+e_i$, with $p\geq 1$,
where $\{e_i\}$ is an i.i.d. mean-zero sequence, with variance $\sigma_e^2<\infty$ and $e_i \sim F$.  The conditional distribution of $X_i$ given $\mathcal{F}_{i-1}=\sigma(X_0,X_1,\ldots,X_{i-1}$) is simply a location shift of the cdf, with conditional mean ${\rm E}[X_i \vert \mathcal{F}_{i-1}]=\phi_0+\phi_1 X_{i-1}+\cdots+\phi_p X_{i-p}$. Thus, the conditional quantile function of $X_i$ is 
$Q_{\tau} \left(X_i \mid \mathcal{F}_{i-1}\right)=\phi_0+ \sum_{j=1}^{p} \phi_j X_{i-j}+F^{-1}(\tau) = \alpha(\tau) + \sum_{j=1}^{p} \phi_j X_{i-j}
= \boldsymbol{U}_i^{\top} \boldsymbol{\theta}(\tau),$ 
with $\boldsymbol{\theta}(\tau) = (\alpha(\tau), \phi_1,\ldots,\phi_p)^{\top}$,  $\alpha(\tau)= \phi_0 + F^{-1}(\tau)$ and $\boldsymbol{U}_i = (1, X_{i-1},\ldots,X_{i-p})^{\top}$.
  Generalizing (\ref{QAR_stat_case}) to the AR($p$) case, we  write
$Q_{\tau}\left(X_i \mid \mathcal{F}_{i-1}\right)=\boldsymbol{U}_i^{\top} \boldsymbol{\theta}(\tau).$  While the conditional mean ${\rm E}[X_i \mid \mathcal{F}_{i-1}]$ provides a model for the average only, the conditional quantile $Q_{\tau}\left(X_i \mid \mathcal{F}_{i-1}\right)$ can 
capture different characteristics of the distribution of $X_i$ by specifying different probability levels (quantiles) $\tau$. For example, setting $\tau = 0.5$ we study the median,  while $\tau = 0.15$ concerns the left tail of the distribution.
Adding noise to  $Q_{\tau}\left(X_i \mid \mathcal{F}_{i-1}\right)$, we obtain 
\begin{equation}
X_i= Q_{\tau}\left(X_i \mid \mathcal{F}_{i-1}\right) + e_i(\tau) 
= \boldsymbol{U}_i^{\top} \boldsymbol{\theta}(\tau)  + e_i(\tau), 
\label{Eq. QReg}
\end{equation}
where, for identifiability of  $\alpha(\tau)$, we set that $e_i(\tau)$ has zero $\tau$th quantile. At the cost of a more cumbersome notation, our modeling can be extended to the case where the AR($p$) process includes also some exogenous covariates. However, for the ease of exposition, we do not pursue that model.\\


\textit{Estimation method.} Given observations $\left\{X_i\right\}_{i=1}^n$ of the stationary process,  the vector $\boldsymbol{\theta}(\tau)$ can be estimated by the quantile regression and the autoregression quantile is
$
\hat{\boldsymbol{\theta}}(\tau)=\underset{\boldsymbol{\theta} \in \Theta}{\operatorname{argmin}} \sum_i \rho_\tau\left(X_i-\boldsymbol{U}_i^{\top} \boldsymbol{\theta}(\tau)\right).
$
If the sequence $\left\{e_i\right\}$ contains i.i.d. random variables such, for that each $e_i$, $F$ has a continuous density $f$ with $f(e)>0$ on $\mathcal{E}=\{e: 0<F(e)<1\}$, then  $\hat{\boldsymbol{\theta}}(\tau)$ satisfies 
$\sqrt{n}(\hat{\boldsymbol{\theta}}(\tau)-\boldsymbol{\theta}(\tau)) \rightarrow^{\mathcal{D}} \mathcal N\left(0, \boldsymbol{v}(\tau) \right)$,  
with asymptotic variance given by
\begin{equation}
	\boldsymbol{v}(\tau) =  \boldsymbol{\Gamma}^{-1}  \frac{\tau(1-\tau)}{f\left[F^{-1}(\tau)\right]^2},
\label{Asy_S}
\end{equation}
 where $\boldsymbol{\Gamma}= \rmE \left(\boldsymbol{U}_{i} \boldsymbol{U}_{i}^{\top}\right)$ exists and it is non singular. 

\subsubsection{Locally stationary case}

\textit{Key notions}. In what follows, for a generic random variable $Z$, we write $Z\in L_q$  for $q>0$, if its $L_q$ norm $\Vert Z \Vert_q = [{\rm E}(|Z|^q)]^{1/q} < \infty$. 
We  denote by $C^k[0,1]$ the set of functions on $[0, 1]$ with $k$th order continuous derivatives. 

To discuss the nonstationary AR($p$) case,  we assume that the model is similar to (\ref{Eq. QReg}), but  the parameters are quantile-varying and time-varying functions and
 $\varepsilon_i(\tau) \in L_2$  with zero $\tau$th quantile. Thus  we set 
\begin{equation}
    X_i 
    = \boldsymbol{\theta}(i/n\mid \tau )^\top \boldsymbol{U}_i + \varepsilon_i(\tau).
    \label{eq:tvmodel}
\end{equation}
Looking at (\ref{eq:tvmodel}), five comments  are in order. First,  the coefficients and hence the quantile dependence structure vary with time, leading to nonstationarity, with  time-varying conditional quantiles $Q_{\tau}\left(X_i \mid \mathcal{F}_{i-1}\right)=  \boldsymbol{\theta}({i}/{n} \mid \tau )^\top \boldsymbol{U}_i $.  Second, the functional form of the time-varying coefficients is not specified: this feature allows the quantile-specific dependence structure to change over time in a nonparametric way. Third,   the model implies that the conditional quantile of $X_i$ depends on its $p$ most recent values: intuitively, as in the stationary case, distant (in time) data do not influence the current observation. 
Fourth, we emphasize the dependence on $n$ of the time-varying coefficients $\phi_j$s: the use of the rescaled time $i/n$ allows us to develop an inference procedure having a meaningful asymptotic theory. Fifth, one may allow each $\phi_j$ to depend on both $i/n$ and $\tau$: this introduces a flexible model, where the slope coefficients change
with the rescaled time and with the quantile. This opens the door to testing for quantile homogeneity, along the same lines of Section 3.3 in \cite{XSZ22}. In this paper, we do not consider the testing problem.

As it is customary in the literature on local stationarity (see \cite{D12}),
we introduce a stationary Markov process which we use to approximate the nonstationary $\{X_i, 1 \leq i \leq n \}$. Thus, we 
define a process $\{X_i(u), i\geq 1\}$, indexed by $u\in(0,1]$:
\begin{equation}
X_i(u)= 
\boldsymbol{\theta}(u  \mid \tau )^\top \boldsymbol{U}_i(u) + \varepsilon_i(\tau).
\label{Xtilde}
\end{equation}
where $\boldsymbol{U}_i(u)=(1, X_{i-1}(u),\ldots,X_{i-p}(u))^{\top}$.  Comparing (\ref{eq:tvmodel}) with (\ref{Xtilde}), it seems intuitively clear that if $i/n$ is near $u$, then $X_{i}$ and ${X}_{i}(u)$ should be close in some norm, like e.g. the
$L_2$ norm. The degree of closeness should depend on both the rescaling factor $n$ and the deviation $\vert i/n - u\vert
$. Thus, the family of processes defined through~\eqref{Xtilde} should provide, in some sense, a reasonable approximation to the  process~\eqref{eq:tvmodel}. The following definition  formalizes these heuristics.  \\

\textbf{Definition}. The process $\{X_i, 1 \leq i \leq n\}$ is said to be locally stationary, if for each  $ u \in (0, 1]$, there exist a stationary process indexed by 
$\{X_i(u), 1 \leq i \leq n \}$ 
and a constant $C > 0$ such that in the local time window about $i/n$, $X_i$ can be approximated by $X_i(u)$ in the sense that 
\begin{eqnarray}
\sup_{u\in(0,1]} \Vert X_i(u) \Vert_2 < \infty, \quad \Vert X_i - X_i(u)  \Vert_2 \leq C \left(\Big\vert i/n-u \Big\vert + \frac{1}{n} \right), \ i=1,...,n. 
\label{Def_LS}
\end{eqnarray}
Moreover, we assume that the initial values of $X_i$ and $X_i(u)$ satisfy (\ref{Def_LS}) also at $u=0$.
Occasionally, when (\ref{Def_LS}) holds, we say that $\{X_i, 1 \leq i \leq n\}$ has a locally stationary approximation $\{X_i(u), 1 \leq i \leq n \}$ in the $L_2$-norm. With this definition in mind,  the model in (\ref{eq:tvmodel}) is called locally stationary quantile regression autoregressive model. \\

To prove that the nonstationary process satisfying (\ref{eq:tvmodel}) 
admits a locally stationary approximation  satisfying  (\ref{Xtilde}), we introduce the following\\

\textbf{Assumption 1}  
We have that: (i) the $\varepsilon_i(\tau) \in L_2$ are i.i.d.; (ii) the coefficients $\alpha(\cdot \vert \tau),\phi_1(\cdot ),...,\phi_p(\cdot ) \in 
C^2[0,1]$, $\sup_{u} \vert \alpha(u \vert \tau)\vert< 1 $ and $\sup_{u} \sum_j  \vert \phi_{j} (u ) \vert < 1$.\\

Assumption 1 essentially imposes some regularity conditions (continuity and smoothness) on the model coefficients (which in principle may all change with $\tau$, see \cite{XSZ22}) and it is standard in the literature on
locally stationary processes. Then, we can prove the following
 
\begin{theorem} 
    Suppose Assumption 1 holds. The process $\{X_i, 1 \leq i \leq n\}$ has a locally stationary approximation $\{X_i(u), 1 \leq i \leq n \}$ in $L_2$ norm.
    \label{Th1}
 \end{theorem}

The proof follows along the lines of the proof of Th. 1 in \cite{XSZ22} and it is omitted.  The theorem 
 allows us to define a class of nonparametric estimators for the time-varying model parameters. \\

\textit{Estimation method}. To estimate the time-varying model parameters in (\ref{eq:tvmodel}), 
we introduce the following\\

 \textbf{Assumption 2}
    $\boldsymbol{\theta}(\cdot|\tau)$ is in $C^{k+1}[0,1]$, for $k\geq 1$. \\
  
Then we write a $k$th order of Taylor's approximation of $\boldsymbol{\theta}({i}/{n} \mid \tau )$ around $u$
\begin{equation}
    \boldsymbol{U}_i^{\top} \boldsymbol{\theta}(i/n \mid \tau ) \approx \boldsymbol{U}_i^{\top} \left\{\boldsymbol{\theta}(u \mid \tau ) +  \left(i/n -  u \right) \boldsymbol{\theta}'(u \mid \tau ) + 
    \dots + \left(i/n - u\right)^k \boldsymbol{\theta}^{(k)}(u \mid \tau )/k!\right\}.
    \label{Exp.Taylor}
\end{equation}
Then, we consider a sequence of localized (in time) polynomial quantile regressions, where the 
time localization is achieved by the kernel $K$,  having bandwidth $b_n$  and satisfying \\

\textbf{Assumption 3}
    (i) $K(\cdot)$, with bounded support, is symmetric and continuously differentiable, and $\int_{\mathbb{R}} K(u) d u=1$.
    (ii) $n b_n \rightarrow \infty$ and $n b_n^{2(k + 2)} \rightarrow 0$. \\
 
Setting for convenience $\boldsymbol{\theta}_0({i}/{n} \mid \tau )=\boldsymbol{\theta}({i}/{n} \mid \tau )$, $K_i(u)=K\left\{(i / n-u) / b_n\right\}$, our locally stationary
autoregression quantile 
is the solution to the following time localized  optimization problem ($k$-order polynomial quantile regression)
\begin{equation}
\begin{gathered}
\left(\hat{\boldsymbol{\theta}}_0(u \mid \tau), \dots , \hat{\boldsymbol{\theta}}^{(k)}(u \mid \tau)\right)=
\underset{\boldsymbol{\theta}_0, \dots, \boldsymbol{\theta}_k}{\operatorname{argmin}} \sum_{i=1}^n \rho_\tau\left\{X_i- \sum_{m = 0}^k\left(i/n-u\right)^m 
\frac{\boldsymbol{U}_i^{\top}\boldsymbol{\theta}_m}{m!}\right\} K_{i}(u).
\end{gathered}
\label{eq:argmin1}
\end{equation}
The derivation of the asymptotics of the local polynomial quantile estimator requires some restrictions on
higher-order moments of the innovation terms (needed for the existence of the asymptotic distribution, see e.g. (ii) in the next assumption): \\ 
  
 \textbf{Assumption 4} 
   (i)  $\left\{\varepsilon_i(\tau)\right\}_i$ are i.i.d., and for each $i, \varepsilon_i(\tau)$ is independent of the historical information 
     $\left\{\left(\boldsymbol{U}_j, \boldsymbol{U}_j(u)\right)\right\}_{j \leq i}$. (ii) $ \varepsilon_i(\tau) \in {L}_{4+2 \epsilon}$ for some $\epsilon>0$. (iii) $\left\{\left(X_i(u), \varepsilon_i(\tau)\right)\right\}_i$ is $\alpha$-mixing with mixing 
     coefficients $\alpha_k$ satisfying $\sum_{k=1}^{\infty} \alpha_k^{\epsilon /(2+\epsilon)}<\infty.$ (iv) The density $f_\tau(\cdot)$ of $\varepsilon_i(\tau)$ is bounded and has bounded derivative. (v) The matrix $\boldsymbol{\Gamma}(u)=\rmE\left[\boldsymbol{U}_i(u) \boldsymbol{U}_i(u)^{\top}\right]$ is nonsingular. \\

Looking at Assumption 1-4, we remark that our assumptions are similar to the ones introduced in \cite{XSZ22} and needed to develop the asymptotic theory of 
the  local linear estimator. However, differently from the results available in literature,  we need to include some additional conditions on the higher-order derivatives (see Assumption 2) of the time-varying coefficients, which are needed
to define the local polynomial estimators and their asymptotics. We aim to estimate $\boldsymbol{\theta}(u \mid  \tau)$ at any given time $u \in (0,1)$. Thanks to this set of assumptions,  we achieve this goal complementing the results available in Th. 2 of \cite{XSZ22} and we state

\begin{theorem} \label{Th2}
    Under the Assumptions 1-4, for the $\hat{\boldsymbol{\theta}}_0(u|\tau)$ parameter in (3) in appendix,  we have

    \begin{equation}
        \sqrt{nb_n} \left\{\hat{\boldsymbol{\theta}}_0(u|\tau) - \boldsymbol{\theta}_0(u|\tau) -  b_n^{k+1}  \frac{\boldsymbol{\theta}^{(k+1)}(u\mid \tau)}{(k+1)!} \int_{\mathbb{R}} v^{k+1} K(v) dv\right\}  
        \rightarrow^{\mathcal{D}} \mathcal{N}(0,\boldsymbol{s}(u \mid \tau)), 
        \label{AsyN}
    \end{equation}
with, for $\kappa_2=  \int_{\mathbb{R}} K^2(v) d v$,
    \begin{equation}
			\boldsymbol{s}(u \mid \tau)  =\boldsymbol{\Gamma}(u)^{-1} \frac{\tau(1-\tau)}{f_\tau^2(0)}\kappa_2.
        \label{Asy_LS}
        \end{equation}
    \end{theorem}
    
The proof  makes use of Th. \ref{Th1} and it requires an adaptation of the arguments in \cite{XSZ22}. 
We refer to Appendix 1 (see Supplementary Material)  for the mathematical details. In the following remark, we focus on  some theoretical aspects.\\

\textbf{Remark.} First, the results in \cite{XSZ22} follow as a special case, setting $k=1$ (local linear estimator) 
and without the need for Assumption 2 on higher-order derivatives. Moreover, comparing  (\ref{Asy_LS}) to the asymptotic variance in Eq. (22) of  \cite{XSZ22}, we notice that 
the $k$th order local polynomial estimator has the same asymptotic variability as the local linear and local constant estimator, for $u\in(0,1)$. Second, as in the stationary case, 
Th. \ref{Th2} proves that the asymptotic distribution of the local polynomial estimators is Gaussian and, comparing (\ref{Asy_LS}) to (\ref{Asy_S}), we notice 
that the expression of the asymptotic variance in the locally stationary case is similar to one obtained in the stationary case, but it contains two important differences: the factor  $\kappa_2$ does not exist in (\ref{Asy_S}), since it is related to time localization;  the matrix $\boldsymbol{\Gamma}$ of (\ref{Asy_S}) is replaced by its time localized version $\boldsymbol{\Gamma}(u)$ in  (\ref{Asy_LS}).   Third, (\ref{AsyN}) gives an explicit form 
for the estimator's bias, which is due to nonstationarity. 
The term $\boldsymbol{\theta}^{(k+1)}(u\mid \tau)$ illustrates that the bias depends on the degree of time-variability of the model parameters:  if one believes that the 
 model parameters have a complex time-varying structure,  the use of a higher-order polynomial regression can be helpful to control the asymptotic estimation bias, without affecting the estimator's precision. 
 
 
 \subsubsection{Monte Carlo studies} \label{Sec_MC1}

Theorem \ref{Th2} opens the door to the use of polynomial estimators, with the possibility for the ultimate user of selecting the degree of the polynomial via the specification of $k$
in (\ref{eq:argmin1}). An interesting question is related to the performance that different estimators may have in the AR case with time-varying coefficients. To investigate numerically this aspect, we consider the median ($\tau=0.5$) of an AR(3) with  dynamics 
	$	X_i= \alpha(i/n,\tau) + \phi_1(i/n ) X_{i-1}+ \phi_2(i/n) X_{i-2}+ \phi_3(i/n) X_{i-3} + \varepsilon_{i}(\tau),$ and $\varepsilon_{i}(\tau) \sim \mathcal{N}(0,1),$ with $\tau=0.5$, $\alpha(i/n, \tau)=0$, 
 $\phi_2 (i/n) =  0.2 + 0.2 \sin(18  i/n) +  0.608(i/n) - 0.032((i/n)+1)^3)$
and $\phi_1(\cdot ) = \phi_2(\cdot )/10, \phi_3(\cdot ) = \phi_2(\cdot )/3$. The selected functional form is flexible and it creates a challenging time-varying estimation problem: it combines periodic oscillatory behaviour (encoded in the $\sin$ function) in the slope coefficients with a polynomial growth in time (encoded in the powers of $({i}/{n})$).  Moreover, it ensures that the local approximating process $\{X_i(u)\}$ is stationary, at every $u$. We set the sample size $n=3000$ (similar results, unreported, can be obtained also for smaller sample sizes). We estimate the functional parameter $\boldsymbol{\theta}(u)=(\phi_1(u),\phi_2(u),\phi_3(u))$ as in (\ref{eq:argmin1}), setting $k = 0,1,2$, a choice which yields the local constant ($k=0$), the local linear ($k=1$) and the local quadratic ($k=2$) estimators. For the sake of comparison, we use the same bandwidth $b_n=0.05$ for the three estimators and set the same grid of 100 equidistant points for $u$. 

To comment on the results, we focus on the estimation of the function $\phi_2$---identical comments apply to the other functional parameters $\phi_1$ and $\phi_3$, which are just 
rescaled versions of $\phi_2$. In Figure \ref{Fig.2}, we display the estimated values in the form of functional boxplots, as obtained with 100 Monte Carlo runs; see {\url{https://
github.com/manonflx} for the code. The plots illustrate that the three estimators yield very similar results. 
The main difference that can be noticed is that the local linear and local quadratic estimators have slightly larger variance than the local constant estimator at the
boundaries of the time span (in particular, for $u$ close to zero). We conjecture that this may be due to the sample size; see \cite{YJ97} p. 162. 

\begin{figure}
\begin{center}
 \includegraphics[width = 0.85\textwidth, height = 0.295\textheight]{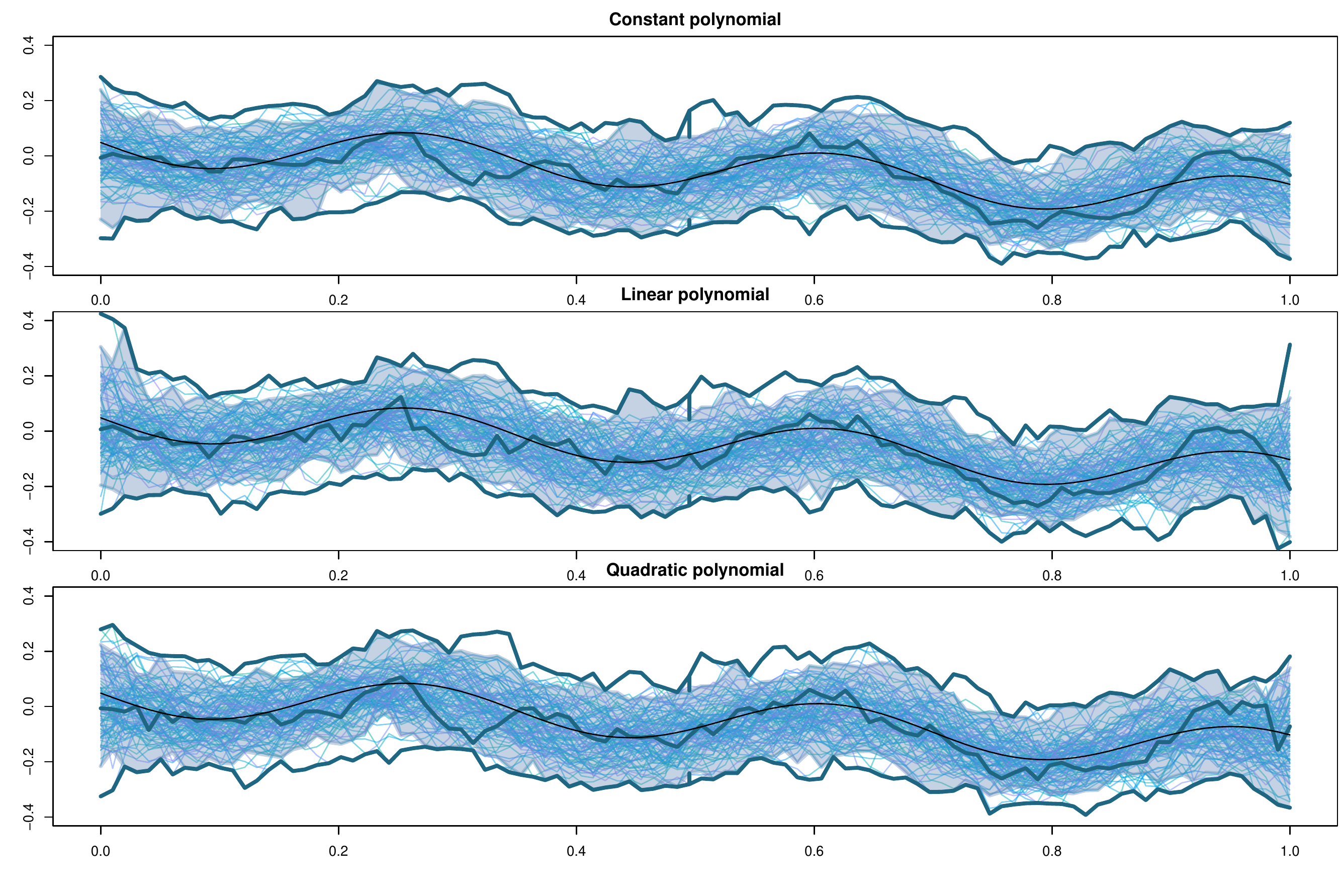}
 \end{center}
 \caption{Functional boxplots of the estimated $\phi_2(u)$, for $\tau=0.5$, with $u\in(0,1)$ (x-axis) as obtained using the local constant, local linear and local quadratic estimation (from top to bottom). In each plot, the middle smooth line represents the true time-varying parameter $\phi_2(u)$}
\label{Fig.2}
 \end{figure}

To have a measure of each estimator performance, we compute the (estimated) MSE at each $u$. In Figure \ref{Fig.3}, we display the MSE curve for each estimator. Remarkably, the three MSEs have very similar values: the three MSE curves are almost overlapping.  Moreover, as it is common to nonparametric kernel estimation procedures, we observe a boundary effect: the MSEs tend to increase when $u$ is close to $0$ or $1$. 
Figure \ref{Fig.2} illustrates that the MSE curves behaviour is principally driven by an increase in the estimators' variance. As far as the computation time is concerned,
we ran our Monte Carlo simulations using and \texttt{Intel(R) core i5-8500 CPU \@ 3.00GHz} and the local constant estimator requires a computation time which is about one and half times smaller than the one requested by the local linear and by the local quadratic estimator, while the local linear is almost as fast as the the local quadratic estimator.  
We performed other simulations (unreported)  with different functional forms of the parameters and our conclusions remain essentially the same. 

In light of this numerical experience, our recommendation is to use the local constant estimator, which offers a performance that is comparable to the performance yielded by the other considered polynomial estimators, but it requires less computational efforts. 


 \begin{figure} 
\begin{center}
 \includegraphics[width = 0.85\textwidth, height = 0.25\textheight]{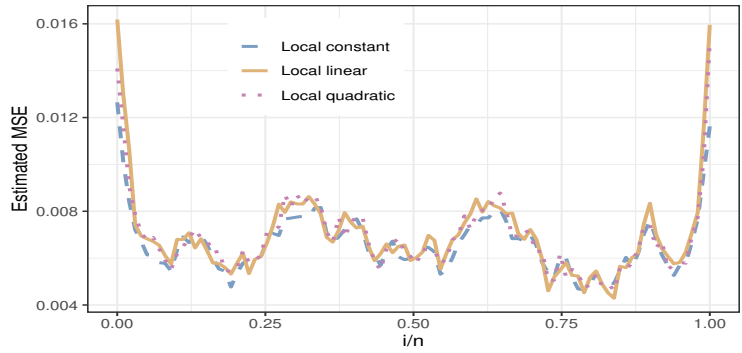}
 \end{center}
 \caption{Estimated MSE at different rescaled time points $u \in (0,1)$ (x-axis) for the local constant, local linear and local quadratic estimator.} 
 \label{Fig.3}
 \end{figure}

\section{Optimal transportation and quantile regression forests}\label{sec:3}

Throughout this section, we consider the setting where $(\X, \Y)$ is an $\mathbb{R}^{m+d}$ random vector with joint distribution ${\rm P}_{\X\Y}$ and marginal distributions ${\rm P}_{\X}$ and ${\rm P}_{\Y}$ for $\X \in \mathbb{R}^{m}$ and $\Y \in \mathbb{R}^{d}$, respectively. Denote by ${\rm P}_{\Y \vert \X = \x}$ the distribution of $\Y$ conditional on $\X = \x$.  Let $(\X\n, \Y\n) := ((\X_1\n, \Y_1\n), \ldots, (\X_n\n, \Y_n\n))$ be a sample of $n$ i.i.d. copies of $(\X, \Y) \sim  {\rm P}_{\X\Y}$. 

\subsection{Basic notions} \label{Bas2}

\subsubsection{Conditional  center-outward quantile, region and contour} \label{Pop}

Based on the measure transportation theory, \cite{Hallin2021} proposed the concept of { center-outward quantile function}, which is the key tool for performing multiple-output quantile regression. To introduce the method, let us denote by $\mathbb S _d$ and ${\bar{\mathbb S}}_d$ the open and closed unit ball, respectively, and by~${\mathcal S}_{d-1}$ the unit hypersphere   in $\mathbb R ^d$. 
Let $\mathcal{P}_d^\pm$ denote the family of all distributions~$\rm P$ with nonvanishing densities $f$, that is,  for all positive $r\in\mathbb{R}$, there exist constants~$L^-_r >0$ and $L^+_r <\infty$ for which   
$L^-_r\leq f({\bf y}) \leq L_r^+$  for all ${\bf y}\in r\,{\bar{\mathbb S}}_d.$
 For $\rm P$ in  this family,  the center-outward distribution and quantile functions defined below  are continuous. 
 Denote by~${\rm U}_d$ the spherical uniform distribution over~${\mathbb S}_d$, that is, the product of a uniform measure over the hypersphere ${\mathcal S}_{d-1}$ and a uniform over the unit interval of distances to the origin.

The {center-outward distribution function}~$\F_{{\pms}}$ of $\rm P$ is defined as the a.e.\ unique gradient of convex function mapping $\mathbb{R}^d$ to $\mathbb{S}_{d}$  
and  { pushing~$\rm P$ forward} to    ${\rm U}_d$ (that is, such that $\F_{{\pms}}({\bf X})\sim{\rm U}_d$ if ${\bf X}\sim{\rm P}$). For ${\rm P}\in{\mathcal P}_d^\pm$, such mapping is  a homeomorphism between~${\mathbb S}_d\setminus\{{\bf 0}\}$ and~$\mathbb{R}^d\setminus \F_{{\pms}}^{-1}(\{{\bf 0}\})$ 
and   the corresponding {center-outward quantile function} is defined  as~$\Q_{\pms} \coloneqq \F_{\pms}^{-1}$ (letting, with a small abuse of notation, $\Q_{\pms} ({\bf 0}) \coloneqq \F_{\pms}^{-1}(\{{\bf 0}\})$). For any given distribution~$\rm P$, the quantile function $\Q_{{\pms}}$ induces a collection of continuous, connected, and nested quantile  contours $\Q_{\pms}(r{\mathcal S}_{d-1})$ and regions $\Q_{\pms}(r{\mathbb S}_{d})$ of order $r\in[0,1)$;   the { center-outward median}  $\Q_{\pms}(\0)$ is a uniquely defined  compact set of Lebesgue measure zero. We refer to \cite{Hallin2021} for details.  These notions allow us to introduce the following \\  

\textbf{Definition} (\cite{dBSH22}) The {conditional  center-outward quantile function} of $\Y$ given $\X$ is the center-outward quantile map ${\bf u} \mapsto \Q_{\pms}({\bf u} \vert \X = {\bf x})$ of ${\rm P}_{\Y \vert \X = \x}, \x \in \mathbb{R}^m$. The corresponding { conditional  center-outward quantile region} and { contour} of order $\tau \in (0, 1)$ are the sets
$\mathbb{C}_{\pms} (\tau \vert \x) := \Q_{\pms}(\tau {\bar{\mathbb S}}_d \vert \X = \x)$  
$\mathcal{C}_{\pms} (\tau \vert \x) := \Q_{\pms}(\tau {{\mathcal S}}_{d-1} \vert \X = \x)$,
respectively.

\subsubsection{Empirical  conditional center-outward quantile}\label{sec.EmpQuantile}

The measure transportation quantities introduced in section \ref{Pop} are at the population level.
Given the sample $(\X\n, \Y\n)$, the empirical version of $\Q_{\pms}({\bf u} \vert \X = {\bf x})$ can be constructed via the two following steps; we refer to \cite{dBSH22}.\\ 

{\it Step 1.} Compute the { empirical conditional distribution} of $\Y$ given $\X = \x$ using the formula
${\rm P}\n_{w(\x)} := \sum_{j = 1}^n w_j\n(\x; \X\n) \delta_{\Y_j},$
where  $\delta_{\Y_j}$ is the Dirac function computed at $\Y_j$ and the sequence of weights $w_j\n, j =1, \ldots, n$ satisfies
$w_j\n(\x; \X\n)  \geq 0 $ and  
$\sum_{j = 1}^n w_j\n(\x; \X\n) = 1.$

{\it Step 2.} Compute the empirical conditional quantiles based on the empirical conditional distribution in Step 1. In order to do this, one needs to first construct a regular grid $\mathfrak{G}^{(N)}$ consisting of $N$ points $\mathfrak{g}_1^{(N)}, \ldots, \mathfrak{g}^{(N)}_N$. To this end\footnote{Other heuristic criteria are possible, see e.g. \cite{HM22}.}, let $N$ factorize into $N=N_R N_S +N_0,$ for~$N_R, N_S, N_0 \in \mathbb{N}$ and~$0\leq N_0 < \min \{ N_R, N_S \}$, where~$N_R \rightarrow \infty$ and $N_S \rightarrow \infty$ as~$N \rightarrow \infty$, and consider a sequence $\mathfrak{G}^{(N)}$ of grids, where each grid consists of the $N_R N_S$ intersections between an $N_S$-tuple $(\boldsymbol{u}_1,\ldots \boldsymbol{u}_{N_S})$ of unit vectors, and the~$N_R$ hyperspheres  with radii $1/(N_R+1),\ldots ,N_R/(N_R+1)$ centered at the origin, along with~$N_0$ copies of the origin. The only requirement is that  the discrete distribution 
${\rm U}_d^{(N)} := {1/N} \sum_{i=1}^N \delta_{\mathfrak{g}_i^{(N)}},$ $N\in \mathbb{N},$
 converges weakly to the uniform~${\rm U}_d$ over the ball $\mathbb{S}_d$. Estimation of the conditional center-outward quantile is based on the optimal transport pushing ${\rm U}_d^{(N)}$ forward to ${\rm P}\n_{w(\x)}$, and it is achieved by solving the linear program
\begin{equation}\label{eq.OT}
\begin{split}
& \quad \min_{\pi := \{\pi_{i, j}\}} \sum_{i=1}^N \sum_{j=1}^n \frac{1}{2} |\Y_j - \mathfrak{g}_i^{(N)} |^2 \pi_{i, j}, \\
& {\rm s.t.} \, \sum_{j=1}^n \pi_{i, j} = \frac{1}{N}, \, i = 1, 2, \ldots, N,\\
& \quad \sum_{i=1}^N \pi_{i, j} =  w_j\n(\x; \X\n), \, j = 1, 2, \ldots, n, \\
& \quad \pi_{i, j} \geq 0,  \,  i = 1, 2, \ldots, N, j = 1, 2, \ldots, n
\end{split},
\end{equation}
with $\vert \cdot \vert$ applied to a vector denoting its Euclidean norm---if applied to a real number, it denotes its absolute value.  Now, let us denote by $\pi^*(\x) = \{\pi^*_{i, j}(\x), i = 1, 2, \ldots, N, j = 1, 2, \ldots, n\}$ the solution of \eqref{eq.OT}. For any gridpoint $\mathfrak{g}^{(N)}_i, i = 1, \ldots, N$, there exists at least one $j \in \{1, 2, \ldots, n\}$ such that $(\mathfrak{g}^{(N)}_i, \Y_j) \in {\rm supp}(\pi^*(\x))$. Since more than one such $j$ may exist, we choose the one which gets the highest mass from $\mathfrak{g}^{(N)}_i$, and in case of ties, we choose the smallest one by letting
$$\Q\n_{w, \pms}(\mathfrak{g}^{(N)}_i \vert \x) := \arg \inf \left\lbrace \vert \y \vert: \y \in {\rm conv}\left( \{Y_J: J \in \arg \max_j  \pi^*_{i, j}(\x)\}\right)\right\rbrace,$$
where ${\rm conv}(\mathcal{A})$ denotes the convex hull of a set $\mathcal{A}$. Supposing the gridpoint $\mathfrak{g}^{(N)}_i$ is on the hypersphere with radius $j/(N_R+1)$, then $\Q\n_{w, \pms}(\mathfrak{g}^{(N)}_i \vert \x)$ is the empirical conditional center-outward quantile of $\Y$ given $\X = \x$ at the level $j/(N_R+1)$.  The corresponding empirical quantile  region $\mathbb{C}\n_{w, \pms} (j/(N_R+1) \vert \x)$ is the set of empirical conditional center-outward quantiles at the level $\tau \leq j/(N_R+1)$, and the empirical quantile contour $\mathcal{C}\n_{w, \pms} (j/(N_R+1) \vert \x)$ is the set of  the quantiles at the level $j/(N_R+1)$. \\


\subsection{Methodology} \label{Meth2}


\subsubsection{A motivating example}

In Step 1 of section~\ref{sec.EmpQuantile}, one needs to specify the weight function 
$w\n: (\x; \X\n) \mapsto  (w_1\n(\x; \X\n), \ldots, w_n\n(\x; \X\n)),$ which is needed to compute ${\rm P}\n_{w(\x)}$. In this section we propose a novel approach for generating weights based on random forests.  The code is available at 
{\url{https://github.com/mayiming24/Empirical-conditional-center-outward-quantile}. To begin with, we use a Monte Carlo simulation to illustrate some issues in the procedure of \cite{dBSH22}, who propose to specify the weights  in Step 1 using a kernel or a $k$-nearest neighbors (kNN) method. In practice, both approaches may encounter some problems; see \cite{bengio2005curse}. 
For instance: (i) they may perform poorly when the data is sparse; (ii) kernel method may not perform well for multidimensional data. To illustrate these points, in Figure~\ref{fig:contour_d}, we plot the empirical conditional center-outward quantile contours at levels $\tau=0.2,0.4,0.6$ with the Gaussian kernel weights (first row) and kNN weights (second row) for $n = 1000$ and $\X \in \mathbb{R}^m$, $m=1, 2, 5$ (the dimension of $\Y$ is $d=2$). We refer to section~\ref{experiments} for details about the data generating process (DGP).  For $m = 1$ (the first column), both types of weights perform reasonably well, with the empirical contours staying close to their population counterparts (the dashed lines).  However, for $m=2$, their performance becomes much worse than the one observed in the $m=1$ case and when $m=5$, we see that the empirical contours yielded by the kernel weights are concentrating at a single point, and those of the kNN at $\tau = 0.4,0.6$ also deviate severely from the population ones.
In the next subsection we illustrate how one can use the random forests to solve these issues.

\begin{figure}[htb]
	\centering
	\begin{subfigure}[b]{0.3\textwidth}
		\centering
		\includegraphics[width=\textwidth]{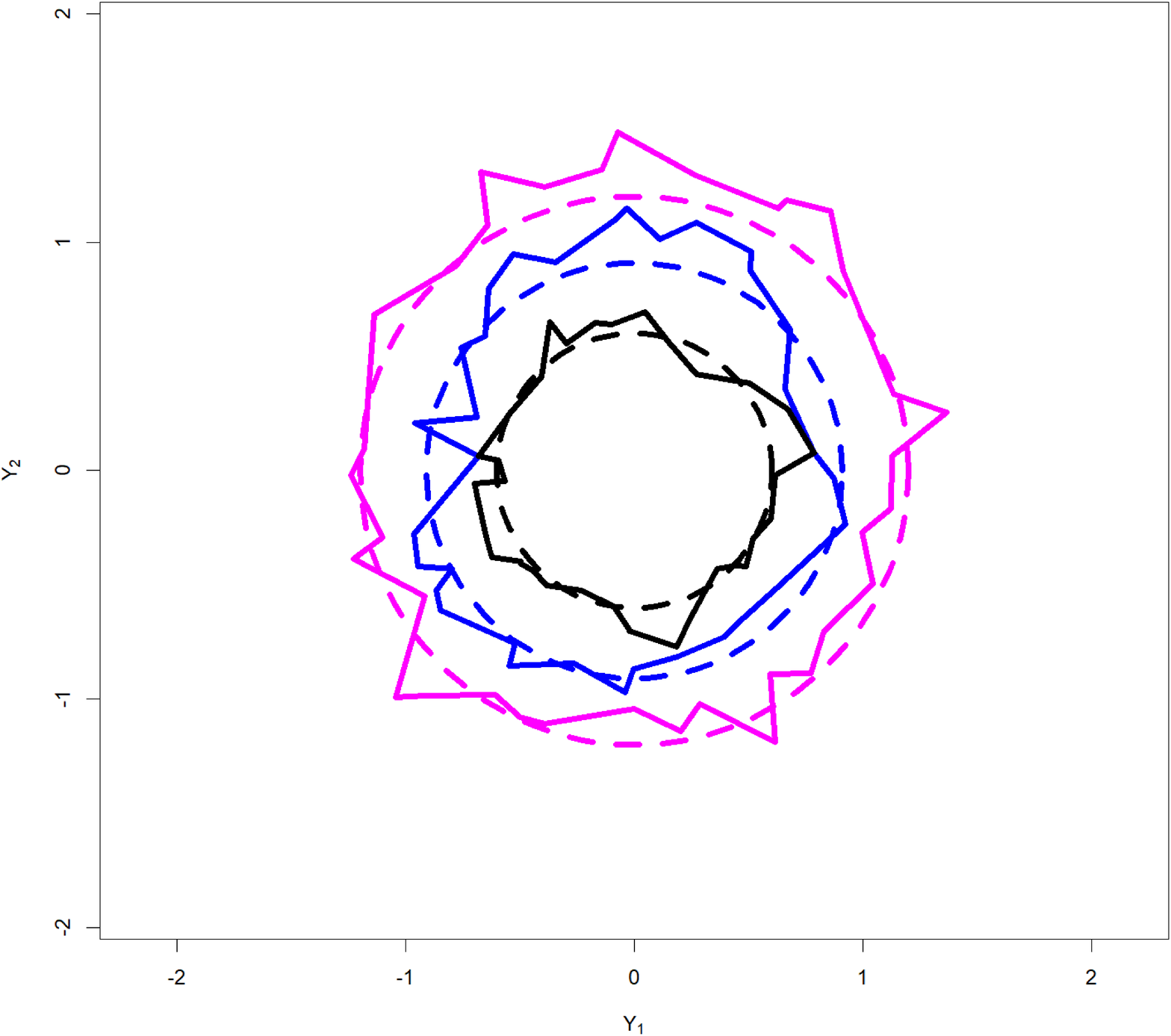}
		\caption{$\mathbf{X} = 0.5 $}
		\label{fig:ck_1}
	\end{subfigure}
	\begin{subfigure}[b]{0.3\textwidth}
		\centering
		\includegraphics[width=\textwidth]{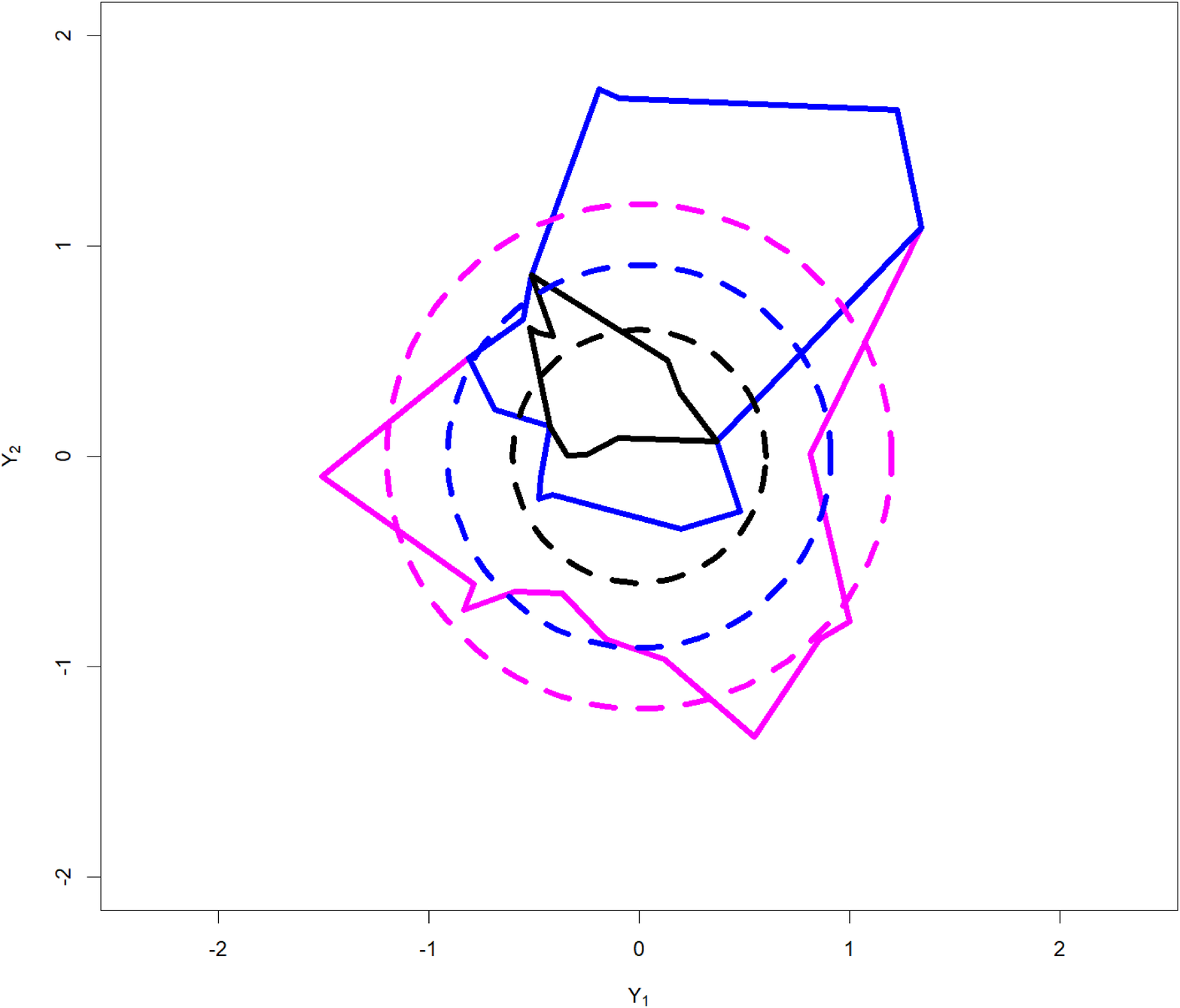}
		\caption{$ \mathbf{X} = (0.5,0) ^{\top}$}
		\label{fig:ck_2}
	\end{subfigure}
	\begin{subfigure}[b]{0.3\textwidth}
		\centering
		\includegraphics[width=\textwidth]{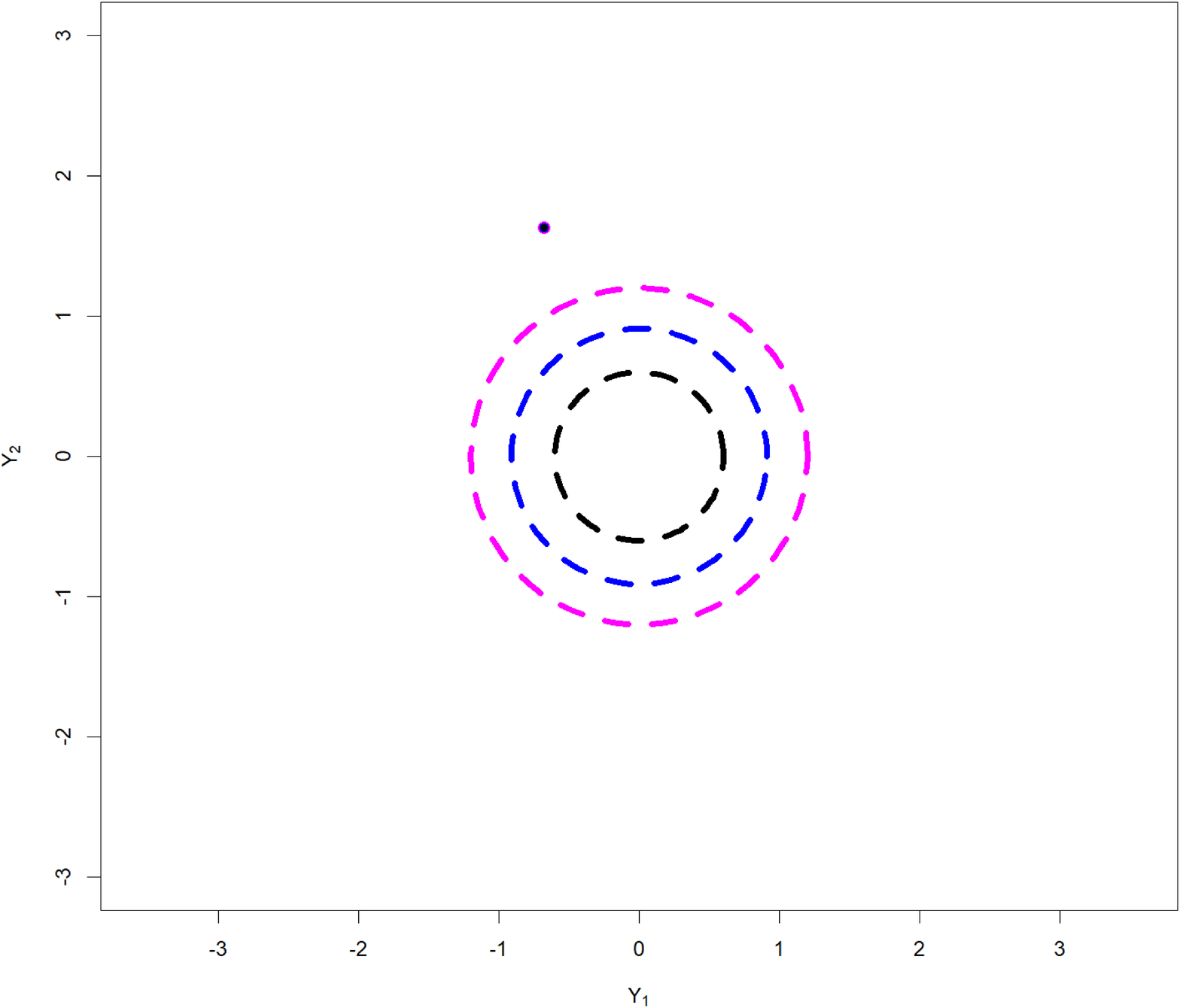}
		\caption{$\mathbf{X}=(0.5,0,0,0,0)^\top$}
		\label{fig:ck_5}
	\end{subfigure}
	\\
	\begin{subfigure}[b]{0.3\textwidth}
		\centering
		\includegraphics[width=\textwidth]{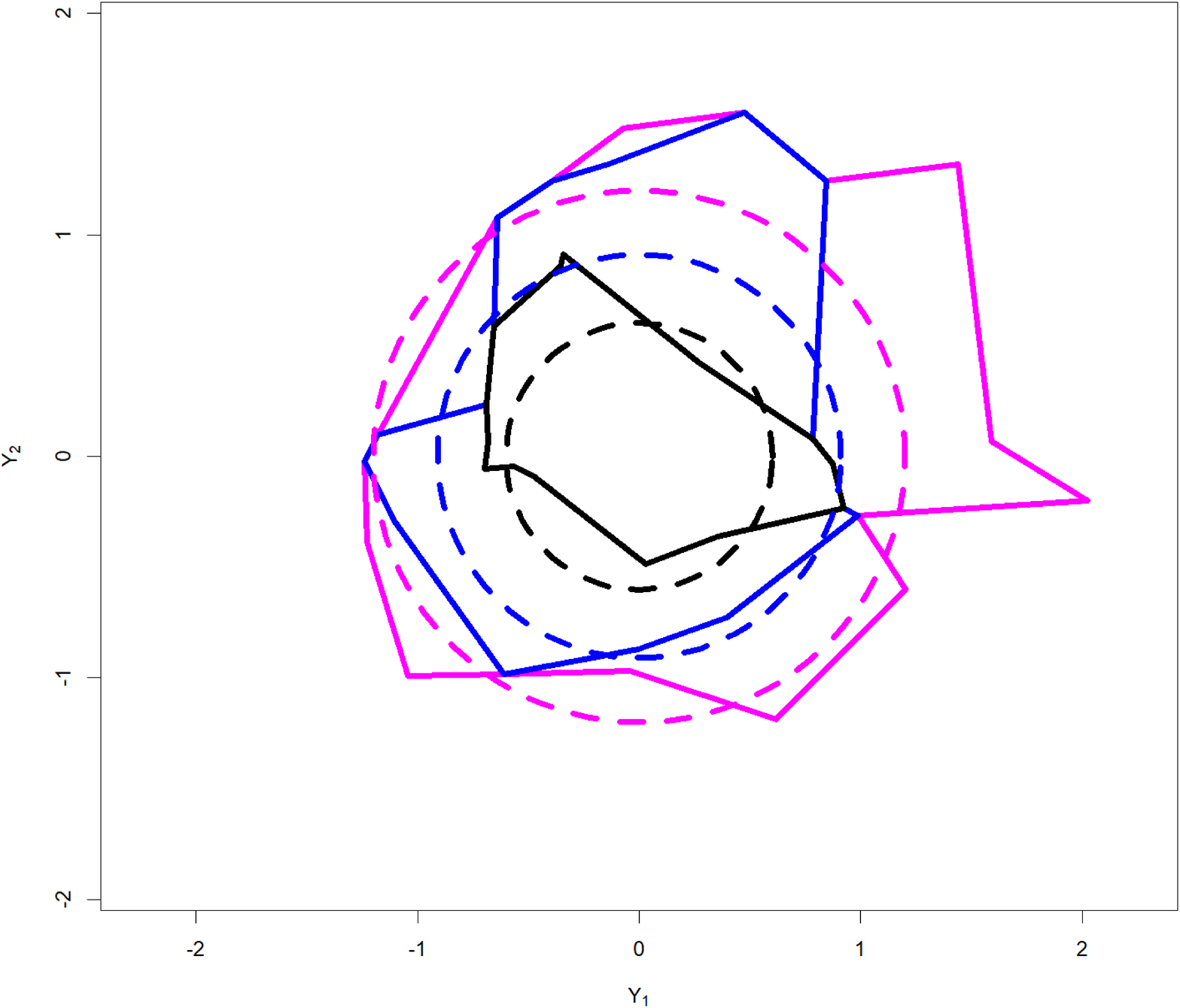}
		\caption{$\mathbf{X} = 0.5 $}
		\label{fig:ckn_1}
	\end{subfigure}
	\begin{subfigure}[b]{0.3\textwidth}
		\centering
		\includegraphics[width=\textwidth]{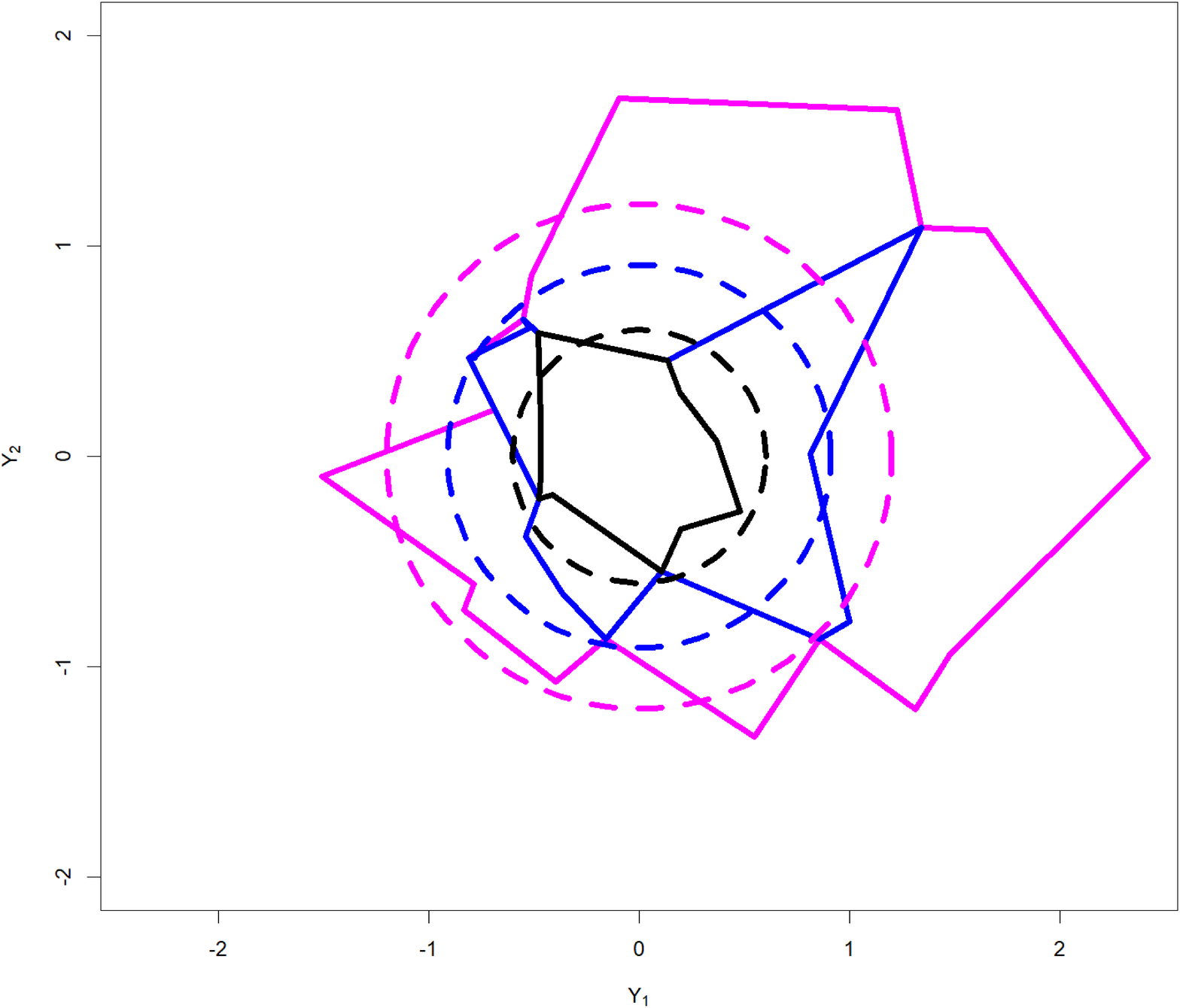}
		\caption{$ \mathbf{X} = (0.5,0)^\top$}
		\label{fig:ckn_2}
	\end{subfigure}
	\begin{subfigure}[b]{0.3\textwidth}
		\centering
		\includegraphics[width=\textwidth]{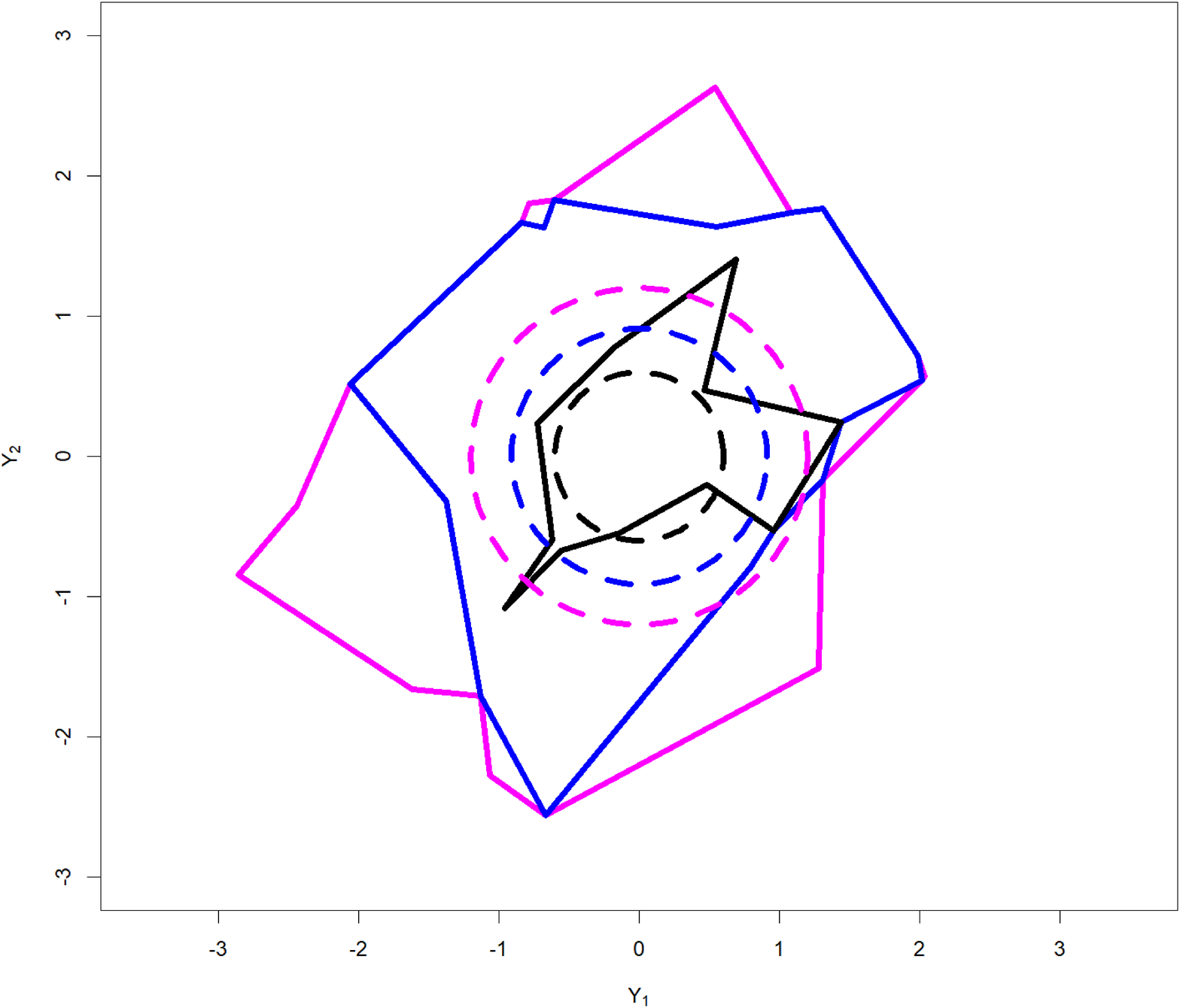}
		\caption{$\mathbf{X}=(0.5,0,0,0,0)^\top$}
		\label{fig:ckn_5}
	\end{subfigure}
	\caption{Plots of the empirical conditional center-outward quantile contours with the kernel weights (first row) and kNN weights (second row). The first, second and third columns are for $m=1, 2, 5$, respectively, and $n=1000$. The  (green,  red  and black) dashed lines represent the   conditional center-outward quantile contours of $\tau=0.2,0.4,0.6$, respectively, and the solid lines are the corresponding empirical contours.} 
	\label{fig:contour_d}
\end{figure}

\subsubsection{Random forests based weights}

To circumvent the problems mentioned above for the kernel and kNN weights, we propose to obtain weights using random forests (henceforth, RF). Following \cite{ATW19}, the weights are generated by averaging neighborhoods produced by different trees. More precisely, we grow a set of $B$ trees indexed by $b = 1, \ldots, B$ and, for each tree, let $L_b(\x)$ denote the set of training examples falling in the same leaf as $\x$. Then the weight $w_j\n(\x; \X\n), j = 1, \ldots, n$ is the averaged (over $B$ trees) frequency that the training sample falls into the same leaf as $\x$, that is,
$w_j\n(\x; \X\n) = {1}/{B} \sum_{b=1}^B w_{bj}\n(\x; \X\n),$
where 
$$w_{bj}\n(\x; \X\n) := \frac{I(\X_j\n \in L_b(\x))}{{\rm card} \{L_b(\x)\}},$$
with ${\rm card} \{L_b(\x)\} $ denoting the number of elements in $L_b(\x)$. Clearly, $w_j\n(\x; \X\n)$ satisfies the conditions mentioned in {\it Step 1.}  by construction. 

We need a criterion for splitting when growing a tree. This criterion is typically based on the minimization of a pre-specified loss function. 
Since the response variable is a random vector in our setting,  taking Euclidean distance between the prediction and test sample in the loss function is not advisable---it ignores the correlation structure between components of the response variable. Therefore, we consider minimizing the Mahalanobis distance between the predictions and observations as in \cite{SX11}.  The use of this distance is dictated by the fact that it has been already implemented in R package {\tt MultivariateRandomForest}. Other distances (e.g. the Wasserstein distance which can be applied to conduct inference beyond the elliptical case, see \cite{HLVL20,hallin2020rank, hallin2023center}) may be considered; this point deserves further  theoretical, methodological and computational investigations. 

\subsection{Monte Carlo studies}\label{experiments}

In this section, we investigate, via Monte Carlo experiments, finite sample performance of the kernel, kNN and RF weights. The DGP is 
\begin{equation}
	\textbf{Y}\n_{i}=  \binom{Y\n_{i_1}}{Y\n_{i_2}} =  (|X^{(n)}_{i_1}|+\ldots+|X^{(n)}_{i_m}| )  \binom{e\n_{i_1}}{e\n_{i_2}}, i = 1, \ldots, n,
\end{equation}
where  $(e\n_{i_1}, e\n_{i_2})^{\top} \sim \mathcal{N}\left((0,  0)^{\top}, \mathbf{I}_2 \right)$ and $ \textbf{X}^{(n)}_i= (X^{(n)}_{i_1},\ldots,X^{(n)}_{i_m} )^{\top}   \sim \mathcal{U}( [-1,1]^m )$. 

To evaluate the performance of different methods, we propose to compare the radius of the population contours with the length of the corresponding empirical quantiles. Note that since $ \textbf{Y}^{(n)}_i$ is sampled from standard normal distribution after scaling, the population conditional quantile contour  $\mathcal{C}_{w, \pm}(\tau \mid \x)$ at each $\x \in \mathbb{R}^m$ is a sphere. Let $R_{\tau}(\x)$  represent the radius of $\mathcal{C}_{w, \pm}(\tau \mid \x)$. So we can evaluate the performance of the method by calculating the distance between each point on the empirical regression quantile contour $ \mathcal{C}_{w, \pm}^{(n)}(\tau \mid \x)$ and the origin (the closer the distance to $ R_{\tau}(\textbf{X}) $, the better the performance). Specifically, assuming that $\mathcal{C}_{w, \pm}^{(n)}(\tau \mid \x)$ has $N_S$ elements, denoted by $\textbf{Y}_1^{\x},\textbf{Y}_2^{\x},\dots,\textbf{Y}_{N_S}^{\x}$,  we define a quantity called MSREC (mean square  radius error of regression quantile contour): 
$
\mathrm{MSREC}_{\tau}(\x) :={1/N_S} \sum_{j=1}^{N_S} (\vert \textbf{Y}_{j}^{\x} \vert - R_{\tau}(\x) )^2.
$ 
Based on the MSREC, we can define the MSRET (mean square  radius error of regression quantile tube) whose intuitive idea is to average MSREC again for the different contours. Note that we should take into account that different $\x$ correspond to different radius, so we consider rescaling by $R^{-2}_{\tau}(\x)$. Suppose there are $N_{\textbf{x}}$ contours, 
$
\mathrm{MSRET}_{\tau} := 1/N_{\textbf{x}} \sum_{k=1}^{N_{\textbf{X}}} {1/R^2_{\tau}(\textbf{x}_k)} \mathrm{MSREC}_{\tau}(\textbf{x}_k).
$

In the following experiments, we set: the bandwidth $b_n=0.1$ in the kernel method; the number of nearest neighbours $k=50$ in the kNN method; the number of  trees  $B= 200$ in RF method---we tried $B = 100, 200, 500, 1000$, and for $B \geq 200$, the results were similar. We apply the Gaussian kernel 
to generate 
$$w_i\n(\x; \X\n) = K\left(\frac{\X_i\n -\x}{b_n}\right) / \sum_{j=1}^n  K\left(\frac{\X_j\n -\x}{b_n}\right), \quad i= 1, \ldots, n.$$

First, we investigate the performance of the three methods from the contour perspective. We set $n=3000$ and $m=2$.
The cross-section plot at $\x = (0.7, 0.7)^\top$ is shown in Figure~\ref{fig:contour}. We find that the empirical conditional center-outward contours of kernel and kNN methods differ significantly from the population contours---e.g. the kNN contours are quite jagged and erratic. On top of that, the estimated contours of different $\tau$-levels overlap: 
this is obviously  undesirable. In contrast, the RF yields better results: the estimated contours have a shape which conforms nicely to the theoretical contours and  yield MSRECs significantly smaller than the others.

\begin{figure}[htb]
	\centering
	\begin{subfigure}[b]{0.32\textwidth}
		\centering
		\includegraphics[width=\textwidth]{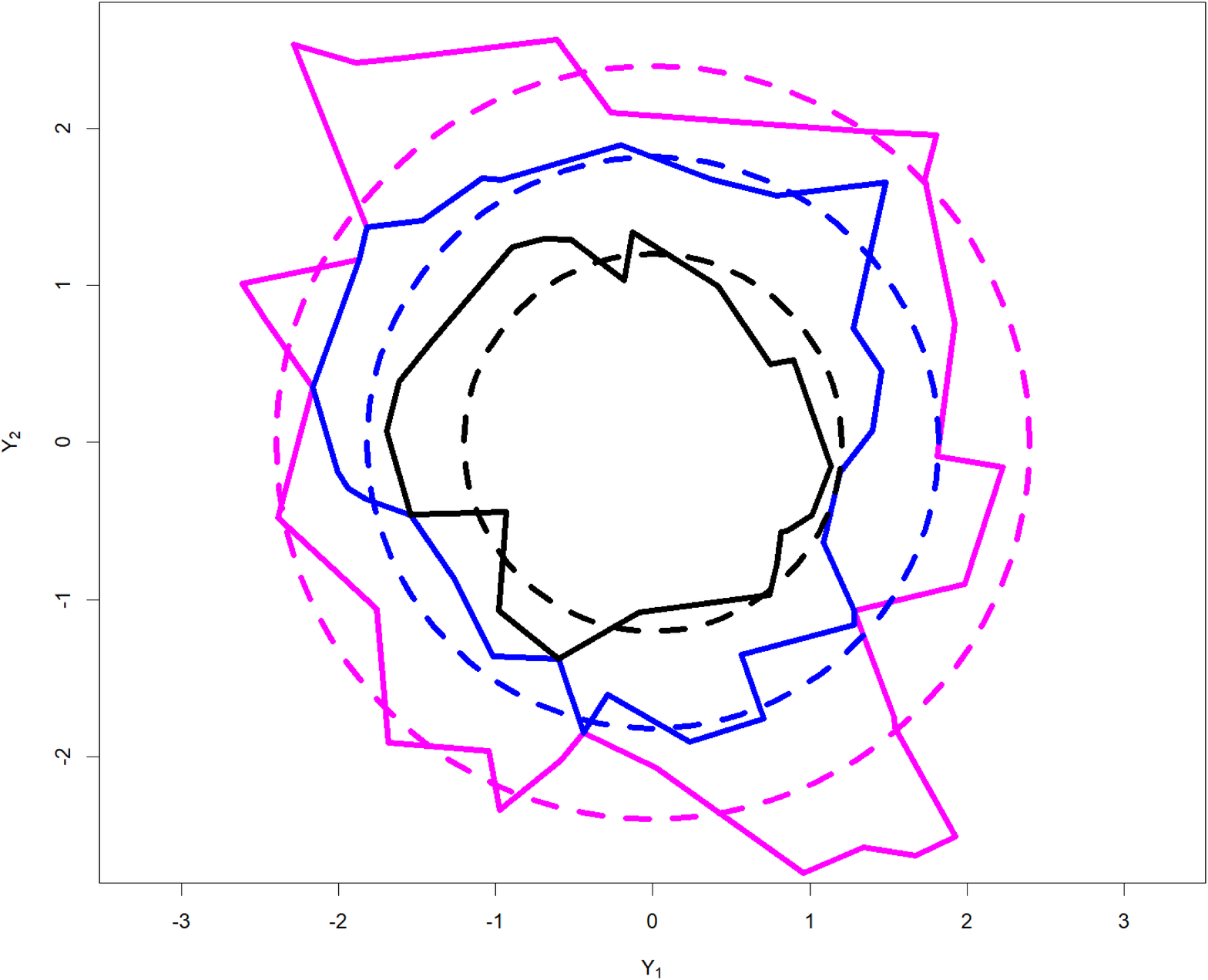}
		\caption{Kernel: $\textrm{MSREC} = 0.0757, 0.0710 ,0.1526$. }
		\label{fig:ck}
	\end{subfigure}
	\begin{subfigure}[b]{0.32\textwidth}
		\centering
		\includegraphics[width=\textwidth]{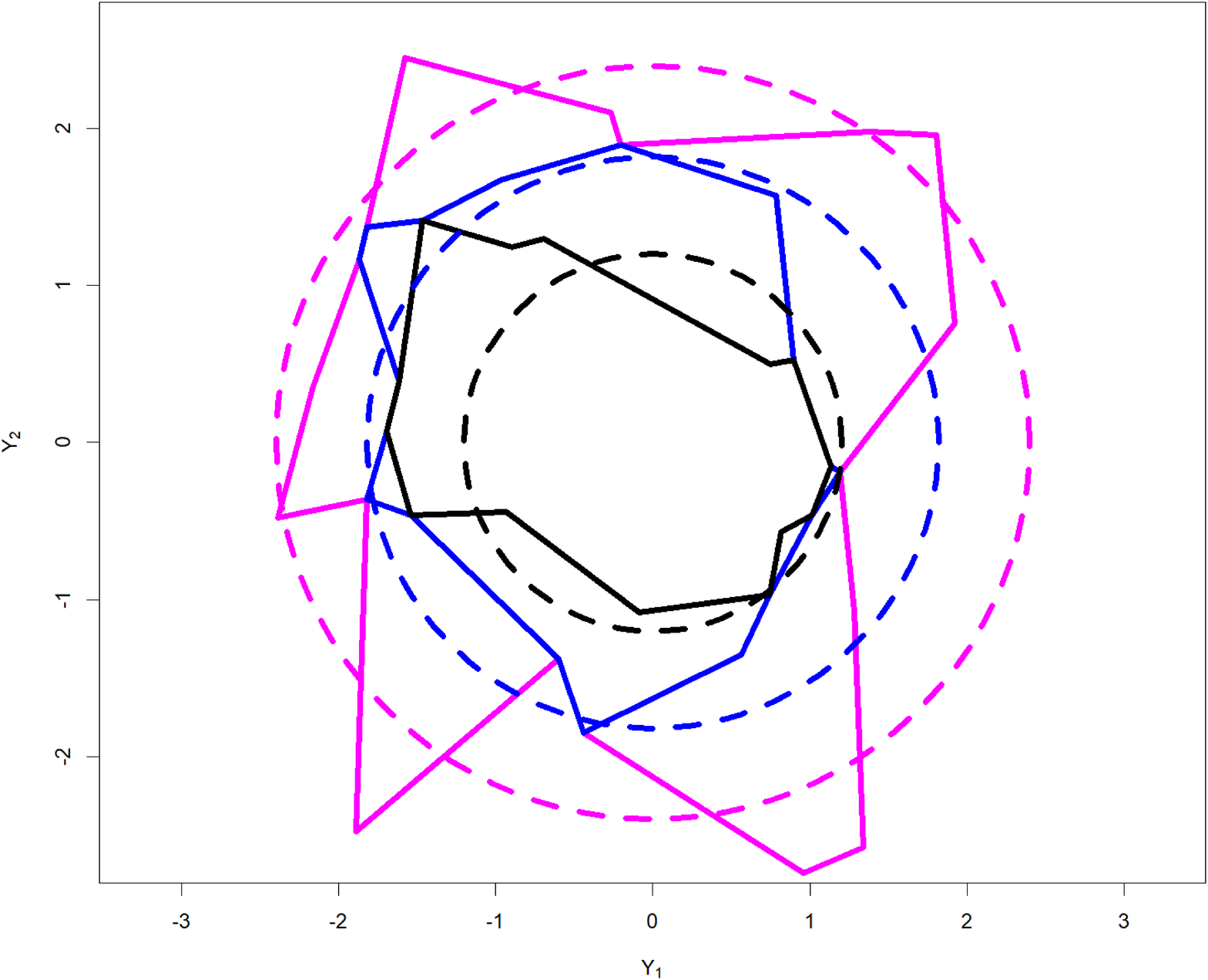}
		\caption{kNN: \textrm{MSREC} = 0.0966, 0.1645, 0.2318.}
		\label{fig:ckn}
	\end{subfigure}
	\begin{subfigure}[b]{0.32\textwidth}
		\centering
		\includegraphics[width=\textwidth]{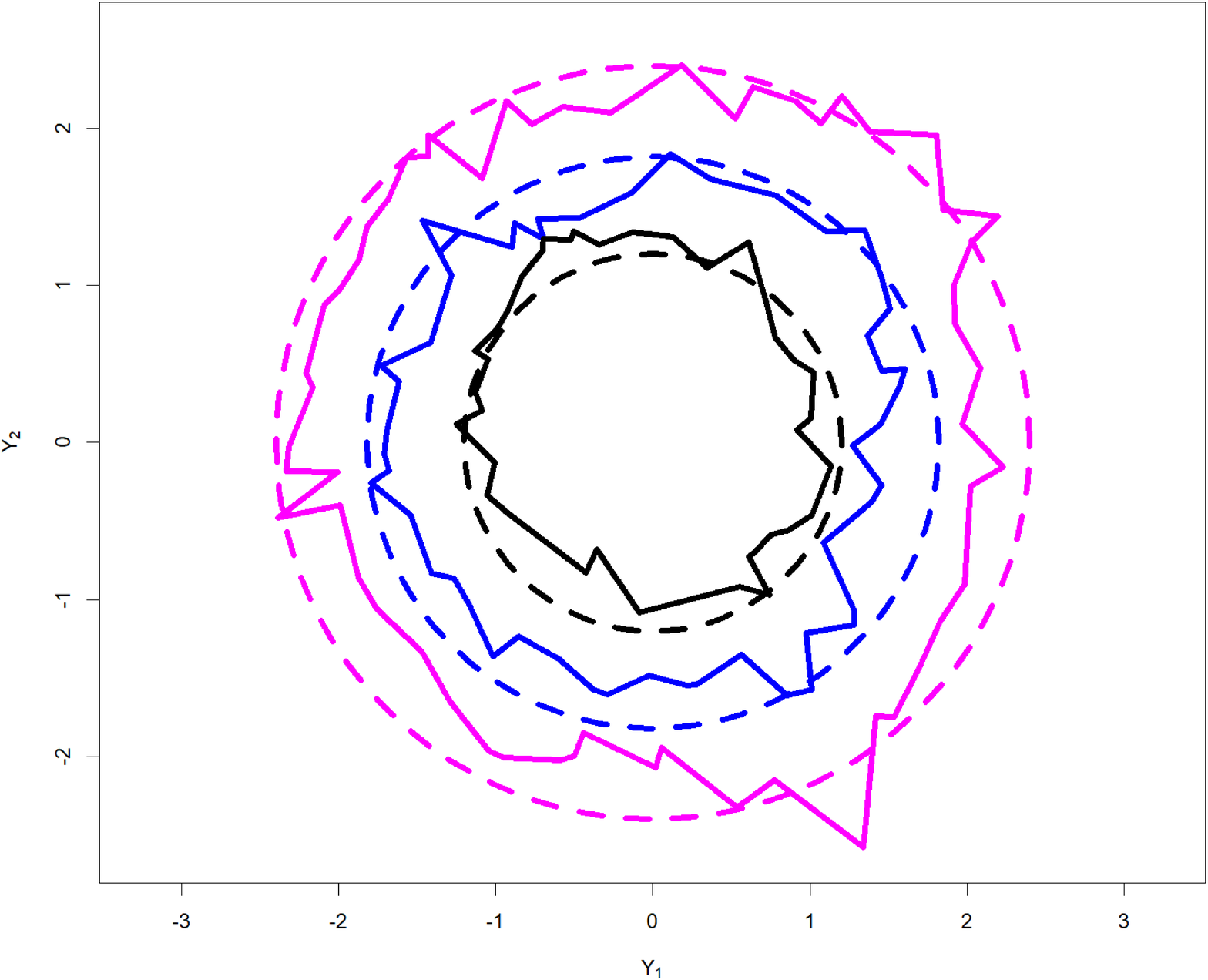}
		\caption{RF: \textrm{MSREC} = 0.0282, 0.0565, 0.0563.}
		\label{fig:cf}
	\end{subfigure}
	\caption{Cross-sectional plot at $\x = (0.7, 0.7)^\top$. The  green,  red  and black dashed lines represent the quantile contours of $\tau=0.2,0.4,0.6$, respectively, and the solid lines are the corresponding empirical results.}
	\label{fig:contour}
\end{figure}

Second, we compute regression quantile tubes. We consider 
(dropping the superscripts for the ease of notation)  
 $ \left\lbrace \x_{j}, j =1,2,...,N_{\x}  \right\rbrace $, where  $N_{\x}=20$,  
$ \x_{j}   = (x_{j_1},0.5)^{\top} $,
and  $\left\lbrace x_{j_1}, j_1=1,2..., N_{\x}\right\rbrace$ is an equally spaced sequence from  -0.9 to 0.9.
Then we  obtain a set of contours, and by considering projection to the three dimensional space containing vectors like  $(x_1, y_1, y_2)^\top$, we get an empirical regression quantile tube. In Figure~\ref{fig:tube} we display the tubes at different $\tau$-levels.  The plots illustrate that over a wide range of $x_1$, the RF method performs best, yielding the smoothest and approximately spherical shaped contours. Also, the \textrm{MSRET} yielded by the RF method (see caption) is much smaller than the one of the other competitors. Information on the computation time of the methods are available in Appendix 2 (see Supplementary Material).

\begin{figure}[htbp]
	\centering
	\begin{subfigure}[b]{0.695\textwidth} 
		\centering
		\includegraphics[width=\textwidth]{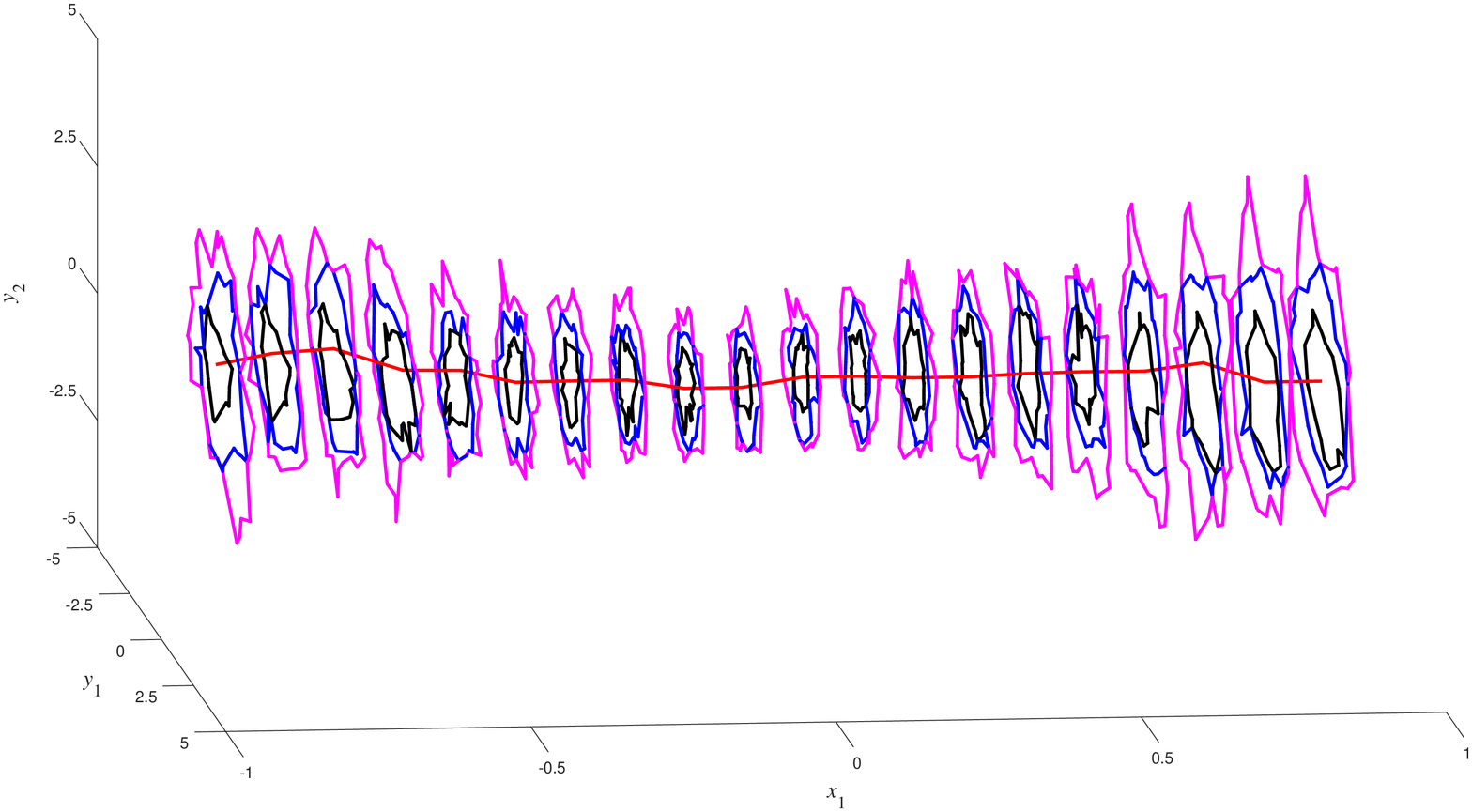}
		\label{fig:tk}
	\end{subfigure}
	\begin{subfigure}[b]{0.695\textwidth}
		\centering
		\includegraphics[width=\textwidth]{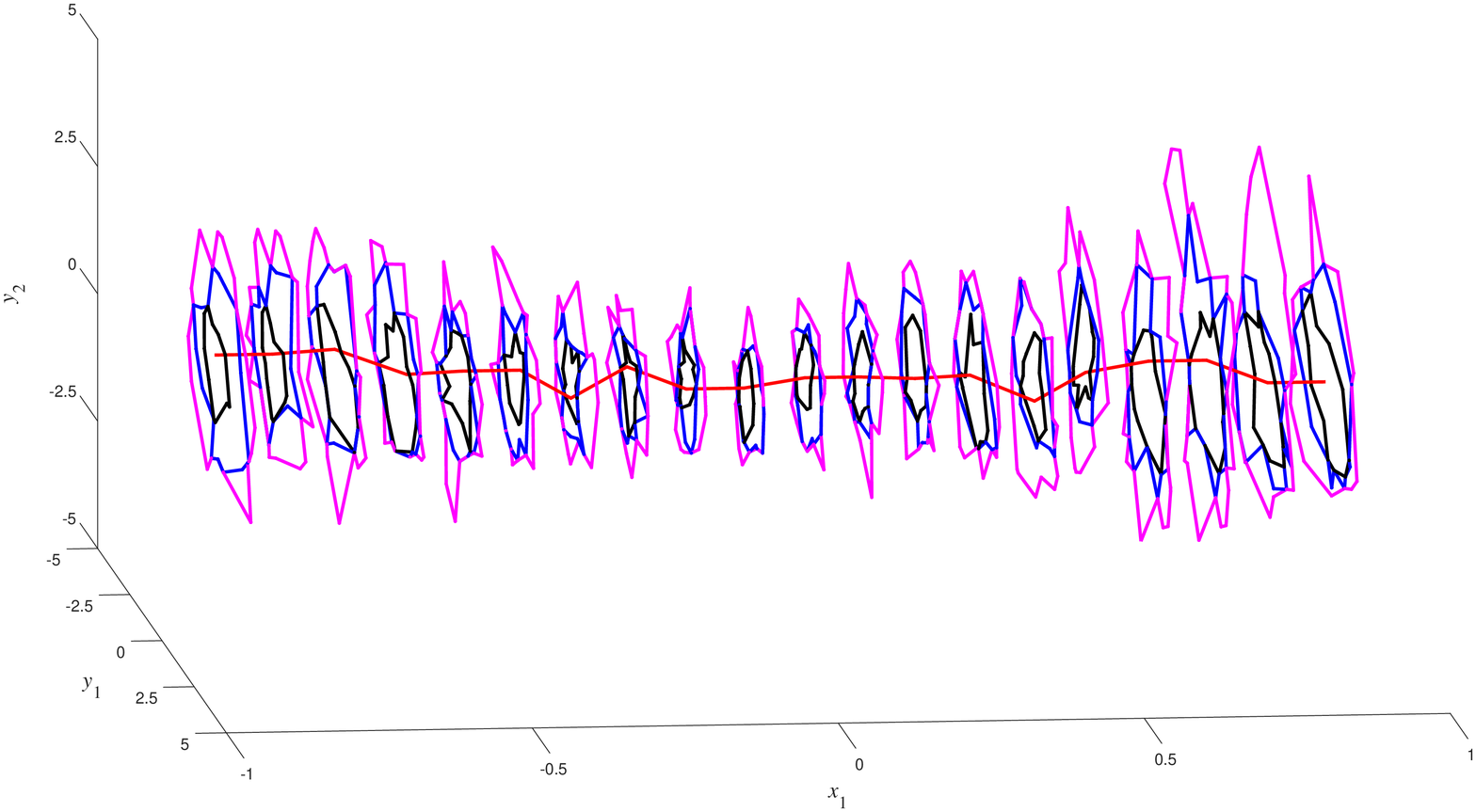}
		\label{fig:tkn}
	\end{subfigure}
	\begin{subfigure}[b]{0.695\textwidth}
		\centering
		\includegraphics[width=\textwidth]{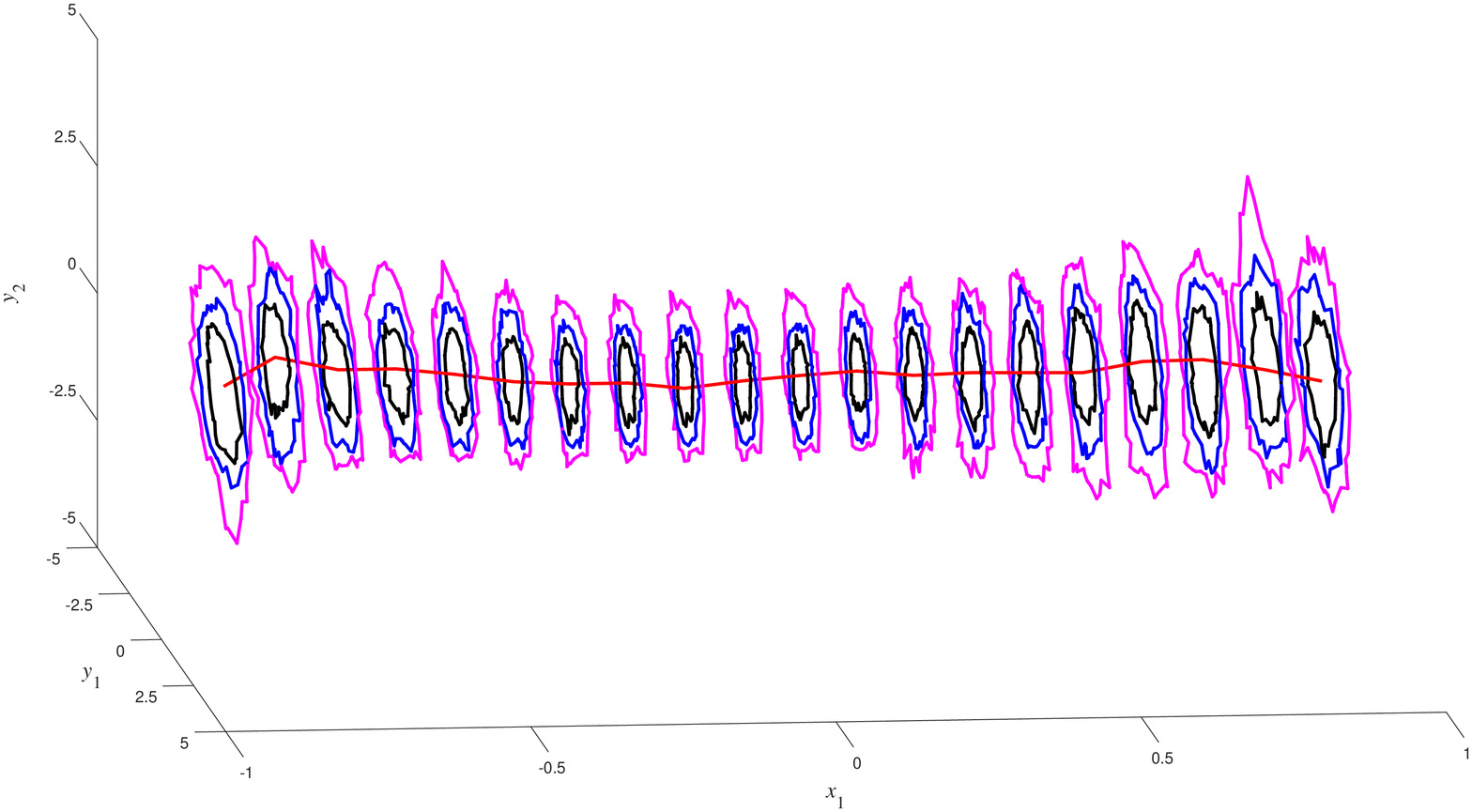}
		\label{fig:tf}
	\end{subfigure}
	\caption{Quantiles tubes at $\tau=0.2,0.4,0.6$ for: Kernel weights (top panel) \textrm{MSRET} = 0.0405, 0.0360, 0.0614; kNN weights (middle panel) \textrm{MSRET} = 0.0601, 0.0718, 0.0929; RF weights (bottom panel) \textrm{MSRET} = 0.0216, 0.0236, 0.0386.   }
	\label{fig:tube}
\end{figure}

Thanks to the adaptive and data-driven property of RF \cite{ATW19, lin2006random}, our method tends to perform better than the kernel and kNN methods in the case of sparse sample distribution. This is often the case when one has to deal with a small sample size and/or when $\mathbf{X}$ has large dimension. We illustrate this aspect keeping the above experimental setup and setting  $n=500,1000,2000,3000$.  $\textrm{MSRET}_{\tau}$ of the kernel, kNN and RF methods  are shown in Table~\ref{tab:samplesize}. For all sample sizes, the RF method outperforms the other methods in terms of $\textrm{MSRET}_{\tau}$.  Notice that the kernel and kNN, when $n=500$, entail very large $\textrm{MSRET}_{\tau}$ compared to the RF. In contrast, $\textrm{MSRET}_{\tau}$ of the RF is stable across $n$.

\begin{table}[htbp]
	\centering
	\begin{tabular}{|c|c|c|c|c|c|c|c|c|c|c|c|c|}
		\hline
		\multirow{2}[4]{*}{} & \multicolumn{3}{c|}{$n=500$} & \multicolumn{3}{c|}{$n=1000$} & \multicolumn{3}{c|}{$n=2000$} & \multicolumn{3}{c|}{$n=3000$} \bigstrut\\
		\cline{2-13}          & Kernel & kNN   & RF & Kernel & kNN   & RF & Kernel & kNN   & RF & Kernel & kNN   & RF \bigstrut\\
		\hline
		$\tau$=0.2 & 0.1937 & 0.0748 & 0.0301 & 0.0783 & 0.0552 & 0.0398 & 0.0569 & 0.0630 & 0.0394 & 0.0405 & 0.0601 & 0.0216 \bigstrut\\
		\hline
	$	\tau$=0.4 & 0.2779 & 0.1287 & 0.0561 & 0.0954 & 0.0608 & 0.0496 & 0.0625 & 0.0723 & 0.0454 & 0.0360 & 0.0718 & 0.0236 \bigstrut\\
		\hline
		$\tau$=0.6 & 0.3808 & 0.1643 & 0.0888 & 0.2237 & 0.1090 & 0.1171 & 0.0828 & 0.1202 & 0.0472 & 0.0614 & 0.0929 & 0.0386 \bigstrut\\
		\hline
	\end{tabular}%
	\caption{\textrm{MSRET} for different sample sizes. }
	\label{tab:samplesize}%
\end{table}%

Now, we consider the case of several dimensions $m$. To this end, we define the $m$-dimensional  $ \left\lbrace \x_{j}, j =1,2,..., N_{\x}\right\rbrace$, where  $N_{\x}=20$,  
$ \x_{j}  = (x_{j_1},0.5, \ldots, 0.5)^{\top} $, and  $\left\lbrace x_{j_1}, j_1=1,2,...,N_{\x}\right\rbrace$ is an equally spaced sequence from -0.9 to 0.9. 
We project to the three dimensional space containing vectors like $(x_1, y_1,y_2)^\top$, and calculate the \textrm{MSRET} of the projection empirical regression quantile tubes. In Table~\ref{tab:dimensions}, we display the results.  In the one-dimensional case ($m=1$),  the numbers in the first three columns  provide evidence of the fact that both the kernel and the kNN method perform well, with a the kernel method yielding the smaller \textrm{MSRET}. The RF method performs similarly to the kNN method, giving similar \textrm{MSRET}, for all choices of $\tau$  and for both sample sizes. In contrast, in the multidimensional cases ($m=2,5$), the RF method provides a clear advantage: its \textrm{MSRET} is smaller than the one entailed by the other methods. To this regard, we emphasize that the RF weights are somewhat similar to the the kNN method weights, in the sense that they are a weighted average of the samples of the nearest neighbours. However, the kNN method fixes  the number of nearest neighbors and their weights to $k$ and ${1}/{k}$, whereas random forests update them adaptively and are hence more flexible. This improves  the \textrm{MSRET} of RF method over the \textrm{MSRET} of the kNN method.

Based on the above numerical evidence, we recommend the use of the RF method for small samples and multidimensional situations.

\begin{table}[htbp]
	\centering
	\begin{tabular}{|c|c|c|c|c|c|c|c|c|c|c|}
		\hline
		\multicolumn{2}{|c|}{\multirow{2}[4]{*}{}} & \multicolumn{3}{c|}{$m=1$} & \multicolumn{3}{c|}{$m=2$} & \multicolumn{3}{c|}{$m=5$} \bigstrut\\
		\cline{3-11}    \multicolumn{2}{|c|}{} & Kernel & kNN   & RF & Kernel & kNN   & RF & Kernel & kNN   & RF \bigstrut\\
		\hline
		\multirow{3}[6]{*}{$n=500$} & $\tau=0.2$ & 0.0373 & 0.0591 & 0.0705 & 0.1937 & 0.0748 & 0.0301 & 0.2445 & 0.0804 & 0.0549 \bigstrut\\
		\cline{2-11}          & $\tau=0.4$ & 0.0438 & 0.0669 & 0.0903 & 0.2779 & 0.1287 & 0.0561 & 0.1426 & 0.0569 & 0.0403 \bigstrut\\
		\cline{2-11}          & $\tau=0.6$ & 0.0779 & 0.1112 & 0.1065 & 0.3808 & 0.1643 & 0.0888 & 0.2666 & 0.1124 & 0.0904 \bigstrut\\
		\hline
		\multirow{3}[6]{*}{$n=1000$} & $\tau=0.2$ & 0.0260 & 0.0656 & 0.0462 & 0.0783 & 0.0552 & 0.0398 & 0.4490 & 0.0500  & 0.0200 \bigstrut\\
		\cline{2-11}          & $\tau=0.4$ & 0.0305 & 0.0633 & 0.0502 & 0.0954 & 0.0608 & 0.0496 & 0.2245 & 0.0894 & 0.0246 \bigstrut\\
		\cline{2-11}          & $\tau=0.6$ & 0.0443 & 0.1053 & 0.0761 & 0.2237 & 0.1090 & 0.1171 & 0.2240 & 0.1384 & 0.0336 \bigstrut\\
		\hline
	\end{tabular}%
	\caption{\textrm{MSRET} for different dimensions of $\x$. }
	\label{tab:dimensions}
\end{table}%

\vspace{-1cm}
\section{Conclusion}

We consider some novel aspects of quantile regression in non standard settings. In the first part of the paper, we focus on AR quantile estimation for locally stationary time series. 
 Beside the  theoretical results discussed here, some open questions remain. For instance, we are planning to study the behaviour of the proposed estimators when $u \rightarrow 0$ and $u \rightarrow 1$ and/or when $\tau \to 1$, in the spirt of extremes. Moreover, similarly to \cite{KS95, HJ99}, we are planning to investigate if and how the notion of autoregression quantiles can yield autoregressive rank scores for testing in the locally stationary setting. 
In the second part of the paper, we explain how to merge generalized random forests with the optimal transportation theory. Our investigation is mainly at the methodological level: the theoretical analysis of our estimation procedure remains the topic for future research. 

\vspace{-0.5cm}
\bibliographystyle{apalike}
\bibliography{HallinPaper}

\begin{thebibliography}{}

\bibitem[Athey et~al., 2019]{ATW19}
Athey, S., Tibshirani, J., and Wager, S. (2019).
\newblock Generalized random forests.
\newblock {\em The Annals of Statistics}, 47(2):1148--1178.

\bibitem[Bengio et~al., 2005]{bengio2005curse}
Bengio, Y., Delalleau, O., and Le~Roux, N. (2005).
\newblock The curse of dimensionality for local kernel machines.
\newblock {\em Techn. Rep}, 1258:12.

\bibitem[Birr et~al., 2017]{BVKDH17}
Birr, S., Volgushev, S., Kley, T., Dette, H., and Hallin, M. (2017).
\newblock Quantile spectral analysis for locally stationary time series.
\newblock {\em Journal of the Royal Statistical Society. Series B (Statistical
  Methodology)}, 79(5):1619--1643.

\bibitem[Breiman, 2001]{B01}
Breiman, L. (2001).
\newblock Random forests.
\newblock {\em Machine learning}, 45(1):5--32.

\bibitem[Dahlhaus, 1997]{D97}
Dahlhaus, R. (1997).
\newblock Fitting time series models to nonstationary processes.
\newblock {\em The Annals of Statistics}, 25(1):1--37.

\bibitem[Dahlhaus, 2012]{D12}
Dahlhaus, R. (2012).
\newblock Locally stationary processes.
\newblock In {Subba Rao}, T., {Subba Rao}, S., and Rao, C., editors, {\em
  Handbook of Statistics}, volume~30, pages 351--412. North Holland.

\bibitem[del Barrio et~al., 2022]{dBSH22}
del Barrio, E., Sanz, A.~G., and Hallin, M. (2022).
\newblock Nonparametric multiple-output center-outward quantile regression.
\newblock {\em arXiv preprint arXiv:2204.11756}.

\bibitem[Del{\'e}amont and La~Vecchia, 2019]{DLV19}
Del{\'e}amont, P.-Y. and La~Vecchia, D. (2019).
\newblock Semiparametric segment {M}-estimation for locally stationary
  diffusions.
\newblock {\em Biometrika}, 106(4):941--956.

\bibitem[El~Bantli and Hallin, 2002]{EH02}
El~Bantli, F. and Hallin, M. (2002).
\newblock Estimation of the innovation quantile density function of an {AR}(p)
  process based on autoregression quantiles.
\newblock {\em Bernoulli}, 8(2):255--274.

\bibitem[Fryzlewicz et~al., 2008]{FSSR08}
Fryzlewicz, P., Sapatinas, T., and Subba~Rao, S. (2008).
\newblock Normalized least-squares estimation in time-varying {ARCH} models.
\newblock {\em The Annals of Statistics}, 36(2):742--786.

\bibitem[Gutenbrunner and Jure{\v{c}}kov{\'a}, 1992]{GJ92}
Gutenbrunner, C. and Jure{\v{c}}kov{\'a}, J. (1992).
\newblock Regression rank scores and regression quantiles.
\newblock {\em The Annals of Statistics}, 20(1):305--330.

\bibitem[Hallin, 2022]{H22}
Hallin, M. (2022).
\newblock Measure transportation and statistical decision theory.
\newblock {\em Annual Review of Statistics and its Application}, 9:401--424.

\bibitem[Hallin et~al., 2021]{Hallin2021}
Hallin, M., del Barrio, E., Cuesta-Albertos, J., and Matr{\'a}n, C. (2021).
\newblock Distribution and quantile functions, ranks and signs in dimension d:
  A measure transportation approach.
\newblock {\em The Annals of Statistics}, 49(2):1139--1165.

\bibitem[Hallin and Jure{\v{c}}kov{\'a}, 1999]{HJ99}
Hallin, M. and Jure{\v{c}}kov{\'a}, J. (1999).
\newblock Optimal tests for autoregressive models based on autoregression rank
  scores.
\newblock {\em The Annals of Statistics}, 27(4):1385--1414.

\bibitem[Hallin et~al., 2007]{hallin2007serial}
Hallin, M., Jure{\v{c}}kov{\'a}, J., and Koul, H.~L. (2007).
\newblock Serial autoregression and regression rank scores statistics.
\newblock In {\em Advances In Statistical Modeling And Inference: Essays in
  Honor of Kjell A Doksum}, pages 335--362. World Scientific.

\bibitem[Hallin et~al., 2020]{HLVL20}
Hallin, M., La~Vecchia, D., and Liu, H. (2020).
\newblock Center-outward {R}-estimation for semiparametric {VARMA} models.
\newblock {\em Journal of the American Statistical Association}, Available on
  line from Dec 2020:1--14.

\bibitem[Hallin et~al., 2022]{hallin2020rank}
Hallin, M., La~Vecchia, D., and Liu, H. (2022).
\newblock Rank-based testing for semiparametric {VAR} models: a measure
  transportation approach.
\newblock {\em Bernoulli}, 29:229--273.

\bibitem[Hallin and Liu, 2023]{hallin2023center}
Hallin, M. and Liu, H. (2023).
\newblock Center-outward rank-and sign-based {VARMA} portmanteau tests:
  {Chitturi}, {Hosking}, and {Li--McLeod} revisited.
\newblock {\em Econometrics and Statistics}.

\bibitem[Hallin and Mordant, 2023]{HM22}
Hallin, M. and Mordant, G. (2023).
\newblock On the finite-sample performance of measure-transportation-based
  multivariate rank test.
\newblock {\em Festschrift for David Tyler,}, pages 87--119.

\bibitem[Hallin and {\v{S}}iman, 2017]{HM17}
Hallin, M. and {\v{S}}iman, M. (2017).
\newblock {\em Multiple-output quantile regression}, volume Handbook of
  {Q}uantile {R}egression, chapter Ch. 12.
\newblock Chapman and Hall/CRC.

\bibitem[Koo and Linton, 2012]{KL12}
Koo, B. and Linton, O. (2012).
\newblock Estimation of semiparametric locally stationary diffusion models.
\newblock {\em Journal of Econometrics}, 170(1):210--233.

\bibitem[Koul and Saleh, 1995]{KS95}
Koul, H.~L. and Saleh, A. M.~E. (1995).
\newblock Autoregression quantiles and related rank-scores processes.
\newblock {\em The Annals of Statistics}, 23(2):670--689.

\bibitem[La~Vecchia et~al., 2023]{LVRI23}
La~Vecchia, D., Ronchetti, E., and Ilievski, A. (2023).
\newblock On some connections between {E}sscher's tilting, saddlepoint
  approximations, and optimal transportation: A statistical perspective.
\newblock {\em Statistical Science}, 38(1):30--51.

\bibitem[Lin and Jeon, 2006]{lin2006random}
Lin, Y. and Jeon, Y. (2006).
\newblock Random forests and adaptive nearest neighbors.
\newblock {\em Journal of the American Statistical Association},
  101(474):578--590.

\bibitem[Meinshausen, 2006]{MR06}
Meinshausen, N. (2006).
\newblock Quantile regression forests.
\newblock {\em Journal of {M}achine {L}earning {R}esearch}, 7(6).

\bibitem[Segal and Xiao, 2011]{SX11}
Segal, M. and Xiao, Y. (2011).
\newblock Multivariate random forests.
\newblock {\em Wiley interdisciplinary reviews: Data mining and knowledge
  discovery}, 1(1):80--87.

\bibitem[Truquet, 2019]{Tr19}
Truquet, L. (2019).
\newblock Local stationarity and time-inhomogeneous {Markov} chains.
\newblock {\em The Annals of Statistics}, 47(4):2023--2050.

\bibitem[Villani, 2009]{V08}
Villani, C. (2009).
\newblock {\em Optimal Transport: Old and New}, volume 338.
\newblock Springer Science \& Business Media.

\bibitem[Vogt, 2012]{V12}
Vogt, M. (2012).
\newblock Nonparametric regression for locally stationary time series.
\newblock {\em The Annals of Statistics}, 40(5):2601--2633.

\bibitem[Xu et~al., 2022]{XSZ22}
Xu, Z., Kim, S., and Zhao, Z. (2022).
\newblock Locally stationary quantile regression for inflation and interest
  rates.
\newblock {\em Journal of Business \& Economic Statistics}, 40(2):838--851.

\bibitem[Yu and Jones, 1997]{YJ97}
Yu, K. and Jones, M. (1997).
\newblock A comparison of local constant and local linear regression quantile
  estimators.
\newblock {\em Computational {S}tatistics \& {D}ata {A}nalysis},
  25(2):159--166.

\bibitem[Zhou and Wu, 2009]{ZW09}
Zhou, Z. and Wu, W.~B. (2009).
\newblock Local linear quantile estimation for nonstationary time series.
\newblock {\em The Annals of Statistics}, 37(5B):2696--2729.

\end{thebibliography}

\newpage

\section{Appendix}

This Appendix contains the supplementary material for the paper having title ``Some novel aspects of quantile regression: local stationarity, random forests and optimal transportation", written for the Festschrift in honour of Professor Marc Hallin. In Section \ref{proof} we provide the proof of Theorem 2, while in Section \ref{App_CompTimes} we give some numerical details about center-outward multi-output quantile regression.

\subsection{Proof of Theorem 2}\label{proof}

In this Appendix, we use the notation $|\boldsymbol z |=\left|z_1\right|+\cdots+\left|z_k\right|$ for a vector $\boldsymbol z=\left(z_1, \ldots, z_k\right)^{\top}$.
For simplicity and readability of the proof, we derive our arguments focusing on an AR(1)---the AR($p$) case follows with some notational changes. Thus,
we consider 
\begin{equation}
	X_i 
	= \boldsymbol{\theta}_0(i/n \mid \tau )^{\top} \boldsymbol{U}_i + \varepsilon_i(\tau),
	\label{eq:model}
\end{equation}
where $ \boldsymbol{\theta}_0(i/n \mid \tau ) = [\alpha(i/n\mid\tau) ,\phi(i/n\mid\tau)]^{\top}$; see \cite{XSZ22} for similar notation.  We denote by $Q_\tau(X_i \mid \mathcal{F}_{i-1})$ the conditional $\tau$th quantile of $X_i$ given its filtration. A $k$th order of Taylor's approximation of $\boldsymbol{\theta}(i/n \mid \tau )$ around $i/n$ yields
\begin{equation}
	\begin{aligned}
		Q_\tau(X_i \mid \mathcal{F}_{i-1}) &= \boldsymbol{U}_i^{\top} \boldsymbol{\theta}_0(i/n \mid \tau ) \\ 
		& \approx \boldsymbol{U}_i^{\top} \left\{\boldsymbol{\theta}_0(u \mid \tau ) +  \left(i/n - u\right)^1 \boldsymbol{\theta}'(u \mid \tau )/1! + \dots + \left(i/n - u\right)^k \boldsymbol{\theta}^{(k)}(u \mid \tau )/k!\right\}.
	\end{aligned}
\end{equation}

We consider the local polynomial quantile regression
\begin{equation}
	\begin{gathered}
		\left(\hat{\boldsymbol{\theta}}_0(u \mid \tau), \hat{\boldsymbol{\theta}}^{\prime}(u \mid \tau), \dots , \hat{\boldsymbol{\boldsymbol{\theta}}}^{(k)}(u \mid \tau)\right)= \\ 
		\underset{\boldsymbol{\theta}_0, \dots, \boldsymbol{\theta}_k}{\operatorname{argmin}} \sum_{i=1}^n \rho_\tau\left\{X_i- \sum_{m = 0}^k\left(i/n-u\right)^m \boldsymbol{U}_i^{\top} \boldsymbol{\theta}_m/m!\right\}K\left(\frac{i/n - u}{b_n}\right) .
	\end{gathered}
	\label{eq:argmin}
\end{equation}

%
To devise the proof for the asymptotics of the resulting estimators, we recall the following two Lemmas from \cite{XSZ22}.
\begin{lemma}
	
	Let $g(\cdot) \in \mathcal{C}^1[0,1]$ be any function. Under Assumption 3, for given $t \in(0,1)$,
	$$
	\sum_{i=1}^n g(i / n)\left(\frac{i/n - u}{b_n}\right)^r K_i(u)=n b_n g(u) \int_{\mathbb{R}} u^r K(u) d u+O\left(n b_n^2+1\right) . \quad r=0,1, \ldots
	$$
\end{lemma}
In particular, by simple calculus, we can show
$$
\sum_{i=1}^n\left(\frac{i/n - u}{b_n}\right)^r K_i(u)=n b_n \int_{\mathbb{R}} u^r K(u) d u+O(1), \quad r=0,1, \ldots
$$
The result of Lemma 1 then easily follows from $g(i / n)=g(u)+O\left(b_n\right)$ for $i/n - u=O\left(b_n\right)$. We will need these calculations for
the proof of Theorem 2. Moreover, we will need also the following
\begin{lemma}
	For each $i$, let $D_i \in {L}_\delta, \delta>2$, be any measurable function of a stationary $\alpha$-mixing process with mixing coefficients $\alpha_k$ satisfying $\sum_{k=1}^{\infty} \alpha_k^{1-2 / \delta}<\infty$. Then
	$$
	\operatorname{var}\left\{\sum_{i=1}^n D_i K_i(u)\right\}=O\left(n b_n \Lambda_n^2\right), \quad \Lambda_n=\max _{1 \leq i \leq n}\left\|D_i\right\|_\delta .
	$$
\end{lemma}

\begin{lemma}
	Consider $\boldsymbol{U}_i$ and its local stationary approximation $\boldsymbol{U}_i(u)$ (see (7) in main text). For any non-random column vectors $\boldsymbol{z}_i \in \mathbb{R}^{p+1}$ that may depend on $n$, define
	\begin{equation*}
		H_n=\sum_{i=1}^n \rho_\tau\left(\boldsymbol{z}_i^{\top} \boldsymbol{U}_i+\varepsilon_i(\tau)\right) K_i(u),
	\end{equation*}
	\begin{equation*}
		\tilde{H}_n=\sum_{i=1}^n \rho_\tau\left(\boldsymbol{z}_i^{\top} \boldsymbol{U}_i(u)+\varepsilon_i(\tau)\right) K_i(u) .
	\end{equation*}
	Suppose Assumption 3 and Assumption 4(i) and (iv) hold. Then
	\begin{equation*}
		H_n-\tilde{H}_n=O_{\rm p}\left[b_n\left(\chi_n+\sqrt{\chi_n}\right)\right] \quad \text { with } \quad \chi_n=\sum_{i=1}^n \boldsymbol{z}_i^{\top} \boldsymbol{z}_i K_i(u) .
	\end{equation*}
	
\end{lemma}
The proofs of these lemmas can be found in the Supplementary Material of \cite{XSZ22}.\\

Now, we can prove the Theorem 2. To start with, let us recall that $\boldsymbol{U}_i = [1, X_{i-1}]^{\top}$ and that  $\boldsymbol{U}_i(u)$ is $\boldsymbol{U}_i$ where we replace the $X_i$ by its stationary approximation $X_i(u)$. The following bounds are valid since the vector $\boldsymbol{U}_i$ is the same as in the proof of \cite{XSZ22}, thus we have

\begin{equation}
	\begin{aligned}
		\rmE\left|\boldsymbol{U}_i-\boldsymbol{U}_i(u)\right| & =O(|i/n - u|+1 / n), \\
		\rmE\left|\boldsymbol{U}_i \left[\boldsymbol{U}_i\right]^{\top}-\boldsymbol{U}_i(u)\left[\boldsymbol{U}_i\right]^{\top}(u)\right| & =O(|i/n - u|+1 / n), \\
		\rmE\left|\left[\boldsymbol{U}_i(u)-\boldsymbol{U}_i\right]\left[\boldsymbol{U}_i(u)-\boldsymbol{U}_i\right]^{\top}\right| & =O\left[(|i/n - u|+1 / n)^2\right] .
	\end{aligned}
\end{equation}
The loss function of the optimization problem is defined by

$$
\mathcal{L}(\boldsymbol{\theta}_0, \dots,\boldsymbol{\theta}_k ) = \sum_{i = 1}^n \rho_\tau \left\{X_i - \sum_{m = 0}^k \boldsymbol{U}_i^{\top} (i/n - u)^m \boldsymbol{\theta}_m\right\} K_i(u).
$$
The vectors $\left(\hat{\boldsymbol{\theta}}_0(u \mid \tau), \hat{\boldsymbol{\theta}^{\prime}}(u \mid \tau), \dots , \hat{\boldsymbol{\theta}}^{(k)}(u \mid \tau)\right)$ are the minimizers of $\mathcal{L}(\boldsymbol{\theta}_0, \boldsymbol{\theta}_1, \dots,\boldsymbol{\theta}_k )$. Define

$$
\boldsymbol{\Delta}_k =\sqrt{n b_n}\left[\begin{array}{c}
	b_n^0 \left\{ \boldsymbol{\theta}_0-\boldsymbol{\theta}_0(u \mid \tau)\right\} \\
	b_n^1\left\{\boldsymbol{\theta}_1-\boldsymbol{\theta}^{\prime}(u \mid \tau)\right\}  \\
	\dots  \\
	b_n^k\left\{\boldsymbol{\theta}_k-\boldsymbol{\theta}^{(k)}(u \mid \tau)\right\} \\
\end{array}\right], \quad  
\boldsymbol{Y}_{i,k}=\left[\begin{array}{c}
	(\frac{i/n - u}{b_n})^0 \boldsymbol{U}_i \\
	(\frac{(i/n - u)}{b_n})^1 \boldsymbol{U}_i \\
	\dots \\
	(\frac{(i/n - u)}{b_n})^k \boldsymbol{U}_i \\
\end{array}\right],
$$
both ${(k+1) \times 1}$ vectors. Note that the covariance matrix of $\boldsymbol{Y}_{i,k}(u)$ has the following form

$$
\rmE\left[\boldsymbol{Y}_{i,k}(u) \boldsymbol{Y}_{i,k}(u)^{\top}\right] = 
\begin{bmatrix}
	\boldsymbol{\Gamma}(u) & (i/n - u) \boldsymbol{\Gamma}(u) & \dots & (i/n - u)^{k} \boldsymbol{\Gamma}(u) \\
	(i/n - u) \boldsymbol{\Gamma}(u) & \dots & \dots & \dots \\
	\dots & \dots & \dots & \dots \\
	(i/n - u)^{k} \boldsymbol{\Gamma}(u) & \dots & \dots & (i/n - u)^{2k} \boldsymbol{\Gamma}(u)
\end{bmatrix} .
$$
We can rewrite 

$$
\begin{aligned}
	\boldsymbol{U}_i^{\top} \delta_{i,k} +\varepsilon_i(\tau)-\frac{\boldsymbol{Y}_{i,k}^{\top} \boldsymbol{\Delta}_k}{\sqrt{n b_n}} = X_i - \sum_{m = 0}^k \boldsymbol{U}_i^{\top} (i/n - u)^m \boldsymbol{\theta}_m
\end{aligned}
$$
as a function of $\boldsymbol{\Delta}_k$ where 

\begin{equation}
	\boldsymbol{\delta}_{i,k} = 
	\begin{cases} 
		\boldsymbol{\theta}_0\left(i/n \mid \tau\right)-\boldsymbol{\theta}_0(u \mid \tau) = O(b_n) & \text{if } k = 0  \\
		\boldsymbol{\theta}_0\left(i/n \mid \tau\right)-\boldsymbol{\theta}_0(u \mid \tau)- \sum_{m = 1}^k \left(i/n-u\right)^m \boldsymbol{\theta}^{(m)}(u \mid \tau)/m! = O(b_n^{k+1})      & \text{if } k \geq 1
	\end{cases}
\end{equation}
and since $\boldsymbol{\theta}(t|\tau) \in C^{k+1}[0,1], \boldsymbol{\delta}_{i,k} = O(b_n^{k+1})$. 
We write the re-parameterized vector
$$
\hat{\boldsymbol{\Delta}}_k=\sqrt{n b_n}\left[\begin{array}{c}
	b_n^0 \left\{\hat{\boldsymbol{\theta}}_0(u \mid \tau)-\boldsymbol{\theta}_0(u \mid \tau) \right\} \\
	\cdots \\
	b_n^k\left\{\hat{\boldsymbol{\theta}}^{(k)}(u \mid \tau)-\boldsymbol{\theta}^{(k)}(u \mid \tau)\right\}
\end{array}\right],
$$
which minimizes the re-parameterized loss

$$\mathcal{L}(\boldsymbol{\Delta}_k):=\sum_{i=1}^n\left(\rho_\tau\left\{\boldsymbol{U}_i^{\top} \boldsymbol{\delta}_{i,k}+\varepsilon_i(\tau)-\frac{\boldsymbol{Y}_{i,k}^{\top} \boldsymbol{\Delta}_k}{\sqrt{n b_n}}\right\}-\rho_\tau \left\{\boldsymbol{U}_i^{\top} \boldsymbol{\delta}_{i,k} +\varepsilon_i(\tau)\right\}\right) K_i(u).$$
We compute a quadratic approximation for $\mathcal{L}(\boldsymbol{\Delta}_k)$ by means of the two following steps. \\

\textbf{1st step.}  The first step shows that the difference between $\mathcal{L}(\boldsymbol{\Delta}_k)$ and its version in which we replace $\boldsymbol{U}_{i}$ by its stationary approximation $\boldsymbol{U}(u)_{i}$ is $\rm o_p(1)$. To develop further, define

$$
\tilde{\mathcal{L}}(\boldsymbol{\Delta}_k):=\sum_{i=1}^n\left(\rho_\tau\left\{\boldsymbol{U}_i(u)^{\top} \boldsymbol{\delta}_{i,k}+\varepsilon_i(\tau)-\frac{\boldsymbol{Y}_{i,k}(u)^{\top} \boldsymbol{\Delta}_k}{\sqrt{n b_n}}\right\}-\rho_\tau\left\{\boldsymbol{U}_i(u)^{\top} \boldsymbol{\delta}_{i,k}+\varepsilon_i(\tau)\right\}\right) K_i(u).
$$
We want to prove that $\mathcal{L}(\boldsymbol{\Delta}_k)-\tilde{\mathcal{L}}(\boldsymbol{\Delta}_k) = o_{\rm p}(1)$ using Lemma 3. Note that 

$$
\begin{aligned}
	\tilde{\mathcal{L}}(\boldsymbol{\Delta}_k)&:=\sum_{i=1}^n\left(\rho_\tau\left\{\boldsymbol{U}_i(u)^{\top} \boldsymbol{\delta}_{i,k}+\varepsilon_i(\tau)-\frac{\boldsymbol{Y}_{i,k}(u)^{\top} \boldsymbol{\Delta}_k}{\sqrt{n b_n}}\right\}-\rho_\tau\left\{\boldsymbol{U}_i(u)^{\top} \boldsymbol{\delta}_{i,k}+\varepsilon_i(\tau)\right\}\right) K_i(u) \\
	& = \sum_{i=1}^n\rho_\tau\left\{\boldsymbol{U}_i(u)^{\top} \boldsymbol{\delta}_{i,k}+\varepsilon_i(\tau)-\frac{\boldsymbol{Y}_{i,k}(u)^{\top} \boldsymbol{\Delta}_k}{\sqrt{n b_n}}\right\}K_i(u) - \sum_{i = 1}^n\rho_\tau\left\{\boldsymbol{U}_i(u)^{\top} \boldsymbol{\delta}_{i,k}+\varepsilon_i(\tau)\right\} K_i(u)
\end{aligned}
$$
where we define the first term of the right-hand side as $\tilde{\mathcal{L}_1}(\boldsymbol{\Delta}_k)$ and the second one as $\tilde{\mathcal{L}}_2$. Now, we know that $\tilde{\mathcal{L}}_2 - \mathcal{L}_2 = O_{\rm p}\left[b_n\left(\chi_{2,n}+\sqrt{\chi_{2,n}}\right)\right]$, where 
$$\chi_{2,n}=\sum_{i=1}^n\left(\boldsymbol{\delta}_{i,k}^{\top} \boldsymbol{\delta}_{i,k}\right) K_i(u) = O(nb_n b_n^{2(k+1)}),$$ 
thus 
$$\tilde{\mathcal{L}}_2 - \mathcal{L}_2 = O_{\rm p}\left[b_n\left(\sqrt{O(nb_nb_n^{2(k+1)})}\right)\right] = o_{\rm p}(1).
$$ 
Moreover, $\tilde{\mathcal{L}_1}(\boldsymbol{\Delta}_k) - \mathcal{L}_1(\boldsymbol{\Delta}_k) = O_{\rm p}\left[b_n\left(\chi_{1,n}+\sqrt{\chi_{1,n}}\right)\right]$ where 
$$
\chi_{1,n}=\sum_{i=1}^n\left(\boldsymbol{\delta}_{i,k}^{\top} \boldsymbol{\delta}_{i,k}+O([nb_n]^{-1})\right) K_i(u) = O(b_n^{2(k+1)} + O([nb_n]^{-1}))O(nb_n).
$$ 
Thus $\mathcal{L}(\boldsymbol{\Delta}_k)=\tilde{\mathcal{L}}(\boldsymbol{\Delta}_k)+o_{\rm p}(1)$ under the assumption that $nb_n^{2(k+1)+1} \rightarrow 0$. \\

\textbf{2nd step.} The second step derives a quadratic approximation to the loss function. By the identity 

\begin{equation}
	\rho_\tau(u-\boldsymbol{\delta})-\rho_\tau(u)=-\boldsymbol{\delta}\left(\tau-\Ind{u<0}\right)+\int_0^{\boldsymbol{\delta}}\left(\Ind{u \leq s}-\Ind{u \leq 0}\right) d s,
\end{equation}
we write

\begin{equation}
	\tilde{\mathcal{L}}(\boldsymbol{\Delta}_k)=-\boldsymbol{A}_{n,k}^{\top} \boldsymbol{\Delta}_k+I_{n,k},
\end{equation}
where
\begin{equation}
	\begin{aligned}
		& \boldsymbol{A}_{n,k}=\frac{1}{\sqrt{n b_n}} \sum_{i=1}^n\left[\tau-\Ind{\boldsymbol{U}_i(u)^{\top} \boldsymbol{\delta}_{i,k}+\varepsilon_i(\tau)<0}\right] K_i(u) Y_i(u), \\
		& I_{n,k}=\sum_{i=1}^n \eta_i K_i(u), \quad \\
		& \eta_i=\int_0^{\frac{\boldsymbol{Y}_{i,k}(u)^{\top} \boldsymbol{\Delta}_k}{\sqrt{n b_n}}}\left[\Ind{\boldsymbol{U}_i(u)^{\top} \boldsymbol{\delta}_{i,k}+\varepsilon_i(\tau) \leq s}-\Ind{\boldsymbol{U}_i(u)^{\top} \boldsymbol{\delta}_{i,k}+\varepsilon_i(\tau) \leq 0}\right] d s . \\
	\end{aligned}
\end{equation}
We start by focusing on $I_{n,k}$. Using the inequality $\left|\int_0^{\boldsymbol{\delta}_k} \left(\Ind{u \leq s}-\Ind{u \leq 0}\right) d s\right| \leq|\boldsymbol{\delta}_k| \Ind{|u| \leq|\boldsymbol{\Delta}_k|}$, we have

$$
\left|\eta_i\right| \leq \frac{\left|\boldsymbol{Y}_{i,k}(u)^{\top} \boldsymbol{\Delta}_k\right|}{\sqrt{n b_n}} \Ind{\left|\boldsymbol{U}_i(u)^{\top} \boldsymbol{\delta}_{i,k}+\varepsilon_i(\tau)\right| \leq\left|\boldsymbol{Y}_{i,k}(u)^{\top} \boldsymbol{\Delta}_k\right| / \sqrt{n b_n}} .
$$
Note that there exists a constant $c_1<\infty$ such that $\left|\boldsymbol{Y}_{i,k}(u)^{\top} \boldsymbol{\Delta}_k\right| \leq c_1\left|\boldsymbol{U}_i(u)\right|$ for a fixed $\boldsymbol{\delta}_k$ which implies that $\left|\boldsymbol{Y}_{i,k}(u)^{\top} \boldsymbol{\Delta}_k\right| = O(\left|\boldsymbol{U}_i(u)\right|)$ for $i/n - u=O\left(b_n\right)$. 
So the bound becomes, 

$$
\left|\eta_i\right| \leq \frac{c_1 |\boldsymbol{U}_i(u)|}{\sqrt{n b_n}} \Ind{\left|\boldsymbol{U}_i(u)^{\top} \boldsymbol{\delta}_{i,k}+\varepsilon_i(\tau)\right| \leq\left|\boldsymbol{Y}_{i,k}(u)^{\top} \boldsymbol{\Delta}_k\right| / \sqrt{n b_n}} .
$$
Also, $\boldsymbol{\delta}_{i,k}=O\left(b_n^{k+1}\right)$ uniformly for $i/n - u=$ $O\left(b_n\right)$. Thus, there exists a constant $c_2$ such that
$$
\left|\eta_i\right| \leq \frac{c_1\left|\boldsymbol{U}_i(u)\right|}{\sqrt{n b_n}} \Ind{\left|\varepsilon_i(\tau)\right| \leq c_2\left|\boldsymbol{U}_i(u)\right|\left(1 / \sqrt{n b_n}+b_n^{k+1}\right)} .
$$
Next we can compute the variance of $I_{n,k}$ by Lemma 2 and the stationarity of $\left\{\left(\boldsymbol{U}_i(u), \varepsilon_i(\tau)\right)\right\}_i$
$$
\begin{aligned}
	\operatorname{var}\left(I_{n,k}\right) &= \operatorname{var}\left(\sum_{i = 1}^n \eta_i K_i\right) \\& \leq n b_n \max _{1 \leq i \leq n}\left\|\frac{c_1\left|\boldsymbol{U}_i(u)\right|}{\sqrt{n b_n}} \Ind{\left|\varepsilon_i(\tau)\right| \leq c_2\left|\boldsymbol{U}_i(u)\right|\left(1 / \sqrt{n b_n}+b_n^{k+1}\right)}\right\|_{2+\epsilon}^2 \\
	& =c_1^2\left\|\left|\boldsymbol{U}_1(u)\right| \Ind{\left|\varepsilon_1(\tau)\right| \leq c_2\left|\boldsymbol{U}_1(u)\right|\left(1 / \sqrt{n b_n}+b_n^{k+1}\right)}\right\|_{2+\epsilon}^2 \rightarrow 0,
\end{aligned}
$$
where the last equality comes from the dominated convergence theorem because of $\left|\boldsymbol{U}_i(u)\right| \in {L}_{2(2+\epsilon)}$ and $\left(n b_n\right)^{-1 / 2}+b_n^{k+1} \rightarrow 0$.

Using the Taylor's expansion 
$$
F_\tau(v)=F_\tau(0)+v f_\tau(0)+O\left(v^2\right),
$$ 
and $\boldsymbol{\delta}_{i,k}=O\left(b_n^{k+1}\right)$ for $i/n - u=$ $O\left(b_n\right)$, with the notation $\kappa_i = \eta_i \mid \boldsymbol{U}_i(u)$, we derive

$$
\begin{aligned}
	\rmE\left[\kappa_i\right]  & =\int_0^{\frac{\boldsymbol{Y}_{i,k}(u)^{\top} \boldsymbol{\Delta}_k}{\sqrt{n b_n}}}\left[F_\tau\left\{s-\boldsymbol{U}_i(u)^{\top} \boldsymbol{\delta}_{i,k}\right\}-F_\tau\left\{-\boldsymbol{U}_i(u)^{\top} \boldsymbol{\delta}_{i,k}\right\}\right] d s \\
	& =\frac{f_\tau(0)\left[\boldsymbol{Y}_{i,k}(u)^{\top} \boldsymbol{\Delta}_k\right]^2}{2 n b_n}+O(1)\left[\frac{\left|\boldsymbol{U}_i(u)\right|^3}{\left(n b_n\right)^{3 / 2}}+\frac{\left|\boldsymbol{U}_i(u)\right| b_n^{2(k+1)}}{\sqrt{n b_n}}\right] \\
	& = \int_0^{\frac{\boldsymbol{Y}_{i,k}(u)^{\top} \boldsymbol{\Delta}_k}{\sqrt{n b_n}}}  [sf_\tau(0) + O((s-\boldsymbol{U}_i(u)^{\top} \boldsymbol{\delta}_{i,k})^2) - O((\boldsymbol{U}_i(u)^{\top} \boldsymbol{\delta}_{i,k})^2)] ds \\ 
	& = \frac{f_\tau(0)\left[\boldsymbol{Y}_{i,k}(u)^{\top} \boldsymbol{\Delta}_k\right]^2}{2 n b_n}+O(1)\left[\frac{\left|\boldsymbol{U}_i(u)\right|^3}{\left(n b_n\right)^{3 / 2}}+\frac{\left|\boldsymbol{U}_i(u)\right|^3 b_n^{2(k+1)}}{\sqrt{n b_n}}\right]
\end{aligned}
$$
and 

$$
\begin{aligned}
	\rmE\left\{\rmE\left[\eta_i \mid \boldsymbol{U}_i(u)\right]\right\} &= \frac{f_\tau(0)}{2 n b_n}\rmE[\left[\boldsymbol{Y}_{i,k}(u)^{\top} \boldsymbol{\Delta}_k\right]^2]  +  O(1)\left[\frac{\rmE\left|\boldsymbol{U}_i(u)\right|^3}{\left(n b_n\right)^{3 / 2}}+\frac{\rmE\left|\boldsymbol{U}_i(u)\right|^3 b_n^{2(k+1)}}{\sqrt{n b_n}}\right] \\
	& =\frac{f_\tau(0)}{2 n b_n} \boldsymbol{\Delta}_k^{\top}\rmE[Y_i(u)\boldsymbol{Y}_{i,k}(u)^{\top}]\boldsymbol{\Delta}_k+  O(1)\left[\frac{1}{\left(n b_n\right)^{3 / 2}}+\frac{b_n^{2(k+1)}}{\sqrt{n b_n}}\right].
\end{aligned}
$$
Thus, using $\left|\boldsymbol{U}_i(u)\right| \in {L}_{2(2+\epsilon)}$ and by Lemma 1, we have

\begin{equation}
	\begin{aligned}
		\rmE\left(I_{n,k}\right) & =\sum_{i=1}^n \rmE\left\{\rmE\left[\eta_i \mid \boldsymbol{U}_i(u)\right]\right\} K_i(u) \\ 
		& =\frac{f_\tau(0)}{2 n b_n} \sum_{i=1}^n \boldsymbol{\Delta}_k^{\top} \rmE\left[Y_i(u) \boldsymbol{Y}_{i,k}(u)^{\top}\right] \boldsymbol{\Delta}_k K_i(u)+O\left(n b_n\right)\left[\frac{1}{\left(n b_n\right)^{3 / 2}}+\frac{b_n^{2(k+1)}}{\sqrt{n b_n}}\right] \\
		& = \frac{f_\tau(0)}{2} \boldsymbol{\Delta}_k^{\top} \boldsymbol{\Omega}(u) \boldsymbol{\Delta}_k + O(1 / \sqrt{nb_n}) + O([n^{-1/2} b_n^{2(k+1)-1/2}) \\
		& \rightarrow \frac{f_\tau(0)}{2} \boldsymbol{\Delta}_k^{\top} \boldsymbol{\Omega}(u) \boldsymbol{\Delta}_k
	\end{aligned}
\end{equation}
where $\boldsymbol{\Omega}(u)=\operatorname{diag}\left\{\boldsymbol{\Gamma}(u), \boldsymbol{\Gamma}(u) \int_{\mathbb{R}} u^{k+1} K(u) d u\right\}$ is a block diagonal matrix. We are therefore able to write the following quadratic approximation 
\begin{equation}
	\tilde{\mathcal{L}}(\boldsymbol{\Delta}_k)=-\boldsymbol{A}_{n,k}^{\top} \boldsymbol{\Delta}_k+I_{n,k} =  -\boldsymbol{A}_{n,k}^{\top} \boldsymbol{\Delta}_k+\frac{f_\tau(0)}{2} \boldsymbol{\Delta}_k^{\top} \boldsymbol{\Omega}(u) \boldsymbol{\Delta}_k+o_{\rm p}(1),
\end{equation}
since $$I_{n,k} - \frac{f_\tau(0)}{2} \boldsymbol{\Delta}_k^{\top} \boldsymbol{\Omega}(u) \boldsymbol{\Delta}_k = o_{\rm p}(1).$$  
The same quadratic approximation also holds for $\mathcal{L}(\boldsymbol{\Delta}_k)$. Therefore by applying  the convexity lemma [Pollard (1991)], $\hat{\boldsymbol{\Delta}}_k$ has the Bahadur representation
$$
\begin{aligned}
	\hat{\boldsymbol{\Delta}}_k & =\underset{\boldsymbol{\Delta}_k}{\operatorname{argmin}}\left\{-\boldsymbol{A}_{n,k}^{\top} \boldsymbol{\Delta}_k+\frac{f_\tau(0)}{2} \boldsymbol{\Delta}_k^{\top} \boldsymbol{\Omega}(u) \boldsymbol{\Delta}_k\right\}+o_{\rm p}(1) \\
	& =\frac{1}{f_\tau(0)} \boldsymbol{\Omega}(u)^{-1} \boldsymbol{A}_{n,k}+o_{\rm p}(1) .
\end{aligned}
$$
From the first component of $\hat{\boldsymbol{\Delta}}_k$, we have the asymptotic Bahadur representation for $\sqrt{n b_n}[\hat{\boldsymbol{\theta}}_0(u \mid \tau)-\boldsymbol{\theta}_0(u \mid \tau)]$

\begin{equation}
	\frac{\boldsymbol{\Gamma}(u)^{-1}}{\sqrt{n b_n} f_\tau(0)} \sum_{i=1}^n\left[\tau-\Ind{\boldsymbol{U}_i(u)^{\top} \boldsymbol{\delta}_{i,k}+\varepsilon_i(\tau)<0}\right] K_i(u) \boldsymbol{U}_i(u)+o_{\rm p}(1)
\end{equation}
Therefore, we have the bias term and stochastic term decomposition:

\begin{equation}
	\sqrt{n b_n}\left[\hat{\boldsymbol{\theta}}_0(u \mid \tau)-\boldsymbol{\theta}_0(u \mid \tau)-\frac{\boldsymbol{\Gamma}(u)^{-1}}{f_\tau(0)} \boldsymbol{B}_{n,k}\right]=\frac{\boldsymbol{\Gamma}(u)^{-1}}{f_\tau(0)} \boldsymbol{W}_n+o_{\rm p}(1),
\end{equation}
where

\begin{equation*}
	\begin{aligned}
		\boldsymbol{W}_n & =\frac{1}{\sqrt{n b_n}} \sum_{i=1}^n\left[\tau-\Ind{\varepsilon_i(\tau)<0}\right] K_i(u) \boldsymbol{U}_i(u), \\
		\boldsymbol{B}_{n,k} & =\frac{1}{n b_n} \sum_{i=1}^n K_i(u) \boldsymbol{\zeta}_i, 
	\end{aligned}
\end{equation*}
with $\boldsymbol{\zeta}_i=\left[\Ind{\varepsilon_i(\tau)<0}-\Ind{\boldsymbol{U}_i(u)^{\top} \boldsymbol{\delta}_{i,k}+\varepsilon_i(\tau)<0}\right] \boldsymbol{U}_i(u) .$ Using the same Taylor's expansion argument as before [remember that $\boldsymbol{\delta}_{i,k}=O\left(b_n^{k+1}\right)$ for $i/n - u=$ $\left.O\left(b_n\right)\right]$, we have 

\begin{equation}
	\begin{aligned}
		\rmE\left[\boldsymbol{\zeta}_i \mid \boldsymbol{U}_i(u)\right] & = \rmE\left[\left[\Ind{\varepsilon_i(\tau)<0}-\Ind{\boldsymbol{U}_i(u)^{\top} \boldsymbol{\delta}_{i,k}+\varepsilon_i(\tau)<0}\right] \boldsymbol{U}_i(u) \mid \boldsymbol{U}_i(u)\right] \\
		& = \boldsymbol{U}_i(u)  \rmE\left[\left[\Ind{\varepsilon_i(\tau)<0}-\Ind{\boldsymbol{U}_i(u)^{\top} \boldsymbol{\delta}_{i,k}+\varepsilon_i(\tau)<0}\right] \mid \boldsymbol{U}_i(u)\right] \\
		& = \boldsymbol{U}_i(u)  [F_\tau(0) - F_\tau(-\boldsymbol{U}_i(u)^{\top} \boldsymbol{\delta}_{i,k})] \\ 
		& = \boldsymbol{U}_i(u) \left[F_\tau(0)  - F_\tau(0) + \boldsymbol{U}_i(u)^{\top} \boldsymbol{\delta}_{i,k} f_\tau(0) + O((\boldsymbol{U}_i(u)^{\top} \boldsymbol{\delta}_{i,k})^2)\right] \\
		& = f_\tau(0) \boldsymbol{U}_i(u) \boldsymbol{U}_i(u)^{\top} \boldsymbol{\delta}_{i,k}+O\left(b_n^{2(k+1)}\right)\left|\boldsymbol{U}_i(u)\right|^3
	\end{aligned}  
\end{equation} 
and $\rmE\left(\boldsymbol{\zeta}_i\right)=\rmE\left\{\rmE\left[\boldsymbol{\zeta}_i \mid \boldsymbol{U}_i(u)\right]\right\}= f_\tau(0) \boldsymbol{\Gamma}(u) \boldsymbol{\delta}_{i,k}+O\left(b_n^{2(k+1)}\right)$. Now, we recall that 

$$
\boldsymbol{\delta}_{i,k} = 
\begin{cases} 
	(i/n - u)\boldsymbol{\theta}'(u \mid \tau) + O(b_n^2) & \text{if } k = 0  \\
	\frac{\left(i/n-u\right)^{k+1}}{(k+1)!} \boldsymbol{\theta}^{(k+1)}(u \mid \tau) + O(b_n^{k+2})    & \text{if } k \geq 1
\end{cases},
$$
thus
\begin{equation*}
	\begin{aligned}
		\frac{\boldsymbol{\Gamma}(u)^{-1}}{f_\tau(0)} \rmE\left(\boldsymbol{B}_{n,k}\right) & = \frac{\boldsymbol{\Gamma}(u)^{-1}}{f_\tau(0)} \frac{1}{nb_n} \sum_{i = 1}^n K_i(u)f_\tau(0) \boldsymbol{\Gamma}(u) \boldsymbol{\delta}_{i,k}+  \frac{1}{nb_n}\sum_{i = 1}^n K_i(u)O\left(b_n^{2(k+1)}\right) \\
		& = \frac{1}{nb_n}\sum_{i = 1}^n K_i(u) \boldsymbol{\delta}_{i,k}+ O(1)O\left(b_n^{2(k+1)}\right) \\
		& = b_n^{k+1}  \boldsymbol{\theta}^{(k+1)}(u \mid \tau) /(k+1)!  \left\{\frac{1}{nb_n}\sum_{i = 1}^n (\frac{i/n - u}{b_n})^{k+1} K_i(u) \right\} + O(b_n^{k+2}) \\
		& =  b_n^{k+1}  \boldsymbol{\theta}^{(k+1)}(u \mid \tau) /(k+1)! \left\{\int u^{k+1} K(u) du + O((nb_n)^{-1})\right\} + O(b_n^{k+2}) \\ 
		& =  b_n^{k+1}  \boldsymbol{\theta}^{(k+1)}(u \mid \tau) /(k+1)! \int u^{k+1} K(u) du + o(([nb_n]^{-1/2})).\\
	\end{aligned}
\end{equation*}

Then, we compute the variance of $\boldsymbol{B}_{n,k}$. Since $\boldsymbol{\delta}_{i,k} = O(b_n^{k+1})$ there exists a constant $c_3$ such that

\begin{equation}
	\rmE\left(\left|\boldsymbol{\zeta}_{i,k}\right|^{2+\epsilon}\right) \leq \rmE\left[\Ind{\left|\varepsilon_i(\tau)\right| \leq c_3 b_n^{k+1}\left|\boldsymbol{U}_i(u)\right|}\left|\boldsymbol{U}_i(u)\right|^{2+\epsilon}\right]=O\left(b_n^{k+1}\right) \rmE\left[\left|\boldsymbol{U}_i(u)\right|^{3+\epsilon}\right],
\end{equation}
which leads to $\max _{1 \leq i \leq n} \rmE\left(\left|\boldsymbol{\zeta}_{i,k}\right|^{2+\epsilon}\right)=$ $O\left(b_n^{k+1}\right)$. Thus, applying Lemma 2 with $\delta=2+\epsilon$, we have that the variance is
\begin{equation}
	\begin{aligned}
		\operatorname{var}[\boldsymbol{B}_{n,k}] & = \frac{1}{n^2b_n^2}\operatorname{var}[\sum_{i=1}^{n}K_i(u)\boldsymbol{\zeta}_i] \\
		& = \frac{1}{n^2b_n^2}O(nb_n\left[\operatorname{max} || \boldsymbol{\zeta}_i ||_{2 + \epsilon}\right]^2) \quad \textit{(by Lemma 3)} \\
		& = O(\frac{1}{n^2b_n^2})O(nb_n\cdot b_n^{2(k+1)}) = o([1/(nb_n)]).
	\end{aligned}
\end{equation}
Therefore, we obtain the following expression for the asymptotic bias 

$$
\frac{\boldsymbol{\Gamma}(u)^{-1}}{f_\tau(0)} \boldsymbol{B}_{n,k} = b_n^{k+1}  \boldsymbol{\theta}^{(k+1)}(u \mid \tau) /(k+1)! \int u^{k+1} K(u) du + o_{\rm p}(([nb_n]^{-1/2})).
$$
Now, regarding the last part of the proof, i.e. a CLT for $\boldsymbol{W}_n$, it remains the same as in \cite{XSZ22} since this term does not depend on $\boldsymbol{\delta}_{i,k}$. This completes the proof.

\subsection{Computation time for different multiple-output quantile regression methods } \label{App_CompTimes}

Prompted by an anonymous Referee, in the next table we report the computation time for  different multiple-output quantile regression methods.

\begin{table}[htbp]
	\centering
	\begin{tabular}{|c|l|l|}
		\hline
		\multicolumn{1}{|l|}{Sample size} & Method & Computation Time  \bigstrut\\
		\hline
		\multirow{4}[8]{*}{$n=500$} & Kernel & 0.232 mins \bigstrut\\
		\cline{2-3}          & KNN   & 0.133 mins \bigstrut\\
		\cline{2-3}          & Random Forest & 1.434 mins \bigstrut\\
		\cline{2-3}          & Random Forest has been trained & 0.223 mins \bigstrut\\
		\hline
		\multirow{4}[8]{*}{$n=1000$} & Kernel & 0.388 mins \bigstrut\\
		\cline{2-3}          & KNN   & 0.162 mins \bigstrut\\
		\cline{2-3}          & Random Forest & 103.574 mins \bigstrut\\
		\cline{2-3}          & Random Forest has been trained & 0.342 mins \bigstrut\\
		\hline
		\multirow{4}[8]{*}{$n=2000$} & Kernel & 0.675 mins \bigstrut\\
		\cline{2-3}          & KNN   & 0.252 mins \bigstrut\\
		\cline{2-3}          & Random Forest & 505.224mins \bigstrut\\
		\cline{2-3}          & Random Forest has been trained & 0.446 mins \bigstrut\\
		\hline
		\multirow{4}[8]{*}{$n=3000$} & Kernel & 1.080 mins \bigstrut\\
		\cline{2-3}          & KNN   & 0.323 mins \bigstrut\\
		\cline{2-3}          & Random Forest & 996.673mins \bigstrut\\
		\cline{2-3}          & Random Forest has been trained & 0.545 mins \bigstrut\\
		\hline
	\end{tabular}%
	\caption{Computation time for different methods, at $\tau=0.6$ and for different sample sizes $n=500, 1000, 2000$ and $3000$.}
	
	\label{tab:addlabel}%
\end{table}%

We see that multivariate random forests are more time consuming than the other two methods. This aspect becomes even more evident when the sample size is large (see e.g. the results for $n=2000$). This is due to the architecture of the routines of the R package \texttt{MultivariateRandomForest}. The main time consuming tasks are the generation of the forest and the computation needed to divide the samples maximizing the  Mahalanobis distance between the sub-nodes. However, in our numerical experiments we noticed that, once the forest is trained, only a small  amount of computation time is needed for the next steps. We recorded the time to calculate quantile tube for $\tau = 0.6$, and we  provide also the time required once  the forest has been trained.  All results are obtained on a computer of \texttt{Intel core i5-12450h}. A possible way to speed up the procedure can be related to use a different (user coded) Python or C++ routine to perform the division of the samples.  

\end{document}